%% file: main.tex
\documentclass[preprint,12pt,]{elsarticle}
\biboptions{sort&compress}
\usepackage[a4paper,top=2cm,bottom=2cm,left=3cm,right=3cm,marginparwidth=1.75cm]{geometry}
\usepackage{amsmath}
\usepackage{amssymb}
\usepackage{amsthm}
\usepackage{bm}
\usepackage{graphicx}
\usepackage[colorlinks=true, allcolors=blue]{hyperref}
\usepackage{xcolor}
\usepackage{subfiles}
\usepackage{subcaption}
\usepackage{comment}
\usepackage{float}

\theoremstyle{remark}

\newcommand{\meanv}[1]{\left\langle#1\right\rangle}
\newcommand{\bxi}{\boldsymbol{\xi}}
\newcommand{\bs}{\boldsymbol{s}}

\newcommand{\bxip}{\boldsymbol{\hat{\xi}}}
\newcommand{\bS}{\bm{\mathcal{S}}}

\newcommand{\EXP}[1]{\exp\left( #1 \right)}

\DeclareMathOperator{\extr}{\mathrm{Extr}}  

\journal{Physica A}

\begin{document}

\begin{frontmatter}



\title{The effect of priors on Learning with Restricted Boltzmann Machines}

\author[label1]{Gianluca Manzan\corref{cor1}\fnref{email1}}
\author[label1]{Daniele Tantari\fnref{email2}}

\fntext[email1]{\texttt{gianluca.manzan2@unibo.it}}
\fntext[email2]{\texttt{daniele.tantari@unibo.it}}

\cortext[cor1]{Corresponding author}

\affiliation[label1]{organization={Department of Mathematics, University of Bologna},
 addressline={Piazza di Porta San Donato 5},
 city={Bologna (BO)},
 postcode={40126},
 country={Italy}
 }


\begin{abstract}
   Restricted Boltzmann Machines (RBMs) are generative models designed to learn from data with a rich underlying structure. In this work, we explore a teacher-student setting where a student RBM learns from examples generated by a teacher RBM, with a focus on the effect of the unit priors on learning efficiency. We consider a parametric class of priors that interpolate between continuous (Gaussian) and binary variables. This approach models various possible choices of visible units, hidden units, and weights for both the teacher and student RBMs.

By analyzing the phase diagram of the posterior distribution in both the Bayes optimal and mismatched regimes, we demonstrate the existence of a triple point that defines the critical dataset size necessary for learning through generalization. The critical size is strongly influenced by the properties of the teacher, and thus the data, but is unaffected by the properties of the student RBM. Nevertheless, a prudent choice of student priors can facilitate training by expanding the so-called signal retrieval region, where the machine generalizes effectively.
    
\end{abstract}


\begin{highlights}
\item A teacher-student unsupervised learning problem analyzed with Restricted Boltzmann Machines  with general priors. 
\item 
Phase diagrams,  derived in the statistical mechanics perspective,  reveal different learning regimes and are sensitive to the choice of priors, ranging from gaussian to binary.

\item 
A critical training set size can be defined as a triple point in the phase diagram, below which learning by generalization becomes impossible.

\item 
Hidden unit priors with Gaussian tails facilitate learning by reducing the critical size or the level of regularization required to enter the generalization phase.

\end{highlights}


\end{frontmatter}



\section{Introduction}
Restricted Boltzmann Machines (RBMs) \cite{barra2012equivalence,mehta2019high, huang2022statistical,decelle2021restricted} are a type of generative stochastic neural network (NN) that can learn a probability distribution over its set of inputs. Their ability to learn rich internal representations \cite{le2008representational,decelle2017spectral, carbone2024fast,bereux2024fast,decelle2024inferring} makes them a fundamental building block in deep learning architectures 
RBMs have a visible layer of $N$ units $\bm{s}= \left\{ s_i \right\}_{1 \leq i \leq N}$, a hidden layer of $P$ units $\bm{\tau} = \left\{ \tau_{\mu} \right\}_{1 \leq \mu \leq P}$ and a set of internal connections $\bxi = \left\{ \xi^{\mu}_i \right\}_{1 \leq i \leq N,1 \leq \mu \leq P}$, which are commonly referred to as weights. Given $\bxi$, the joint distribution of the visible and hidden layers  has the Gibbs structure
\begin{align}\label{eq:Gibbs}
    P \left( \bm{s}, \bm{\tau} \big| \bxi \right) &= z^{-1} \left( \bxi \right) P \left( \bm{s} \right) P \left( \bm{\tau} \right) \exp \left(\sqrt{\frac{\beta}{N}} \sum_{i = 1}^N\sum_{\mu = 1}^P   \xi^{\mu}_i s_i \tau_{\mu}\right),
\end{align}
where  $P \left( \bm{s} \right)$ and $P \left( \bm{\tau} \right)$ are priors on the visible and hidden layers, respectively, $\beta$ is the inverse temperature modulating the weights intensity and $z \left( \bxi \right)$ is the partition function normalizing the distribution. RBMs with given weigths can generate data $\bm{s}$ by sampling the marginal distribution
\begin{equation}
    \label{eq:direct_distribution}
    P \left( \bm{s} \big| \bxi \right) = z^{-1} \left( \bxi \right) \psi \left( \bm{s} ; \bxi \right)= z^{-1} \left( \bxi \right) P(\bm{s}) \mathbb{E}_{\tau} \left[ \exp \left( \sqrt{\frac{\beta}{N}}   \sum_{i,\mu}^{N,P} \xi^{\mu}_i s_i \tau_{\mu}\right) \right]
\end{equation}
where $\mathbb{E}_{\tau}$ is the expectation over the hidden units. Marginal Gibbs distributions of the form (\ref{eq:direct_distribution}) are also known as generalized Hopfield networks 
\cite{barra2018phase,hopfield1982neural,sollich2014extensive,agliari2017neural,agliari2015retrieval,agliari2015hierarchical,agliari2018non}.  
RBMs with generic priors, i.e. the generalized Hopfield networks, have been extensively studied in the statistical mechanics literature, for their properties as models of associative memory \cite{amit1985storing,amit1987statistical} and their challenging connection with the Parisi theory of spin glasses \cite{mezard2009information,barra2015multi, panchenko2015free,decelle2018thermodynamics,barra2023thermodynamics,antenucci2019approximate}. This studies only address the so-called \textit{direct problem}, where the weights are either fixed or randomly sampled from fixed distributions, and the statistical properties of the generative model are investigated.  

In a machine learning context, however, the weights are learned \cite{fischer2014training,salakhutdinov2007restricted,tubiana2018restricted} from a dataset of examples, whose distribution the RBM must reconstruct. Therefore the study 
 of the \textit{inverse problem} of weights optimization is fundamental for a theoretical understanding of the RBM learning mechanisms.
From this perspective, a very useful approach is the so called \textit{teacher-student} setting where a \textit{student} RBM is trained with data produced by another \textit{teacher} RBM \cite{alemanno2023hopfield,huang2016unsupervised,decelle2021inverse, huang2017statistical, barra2017phase, hou2019minimal, theriault2024modelling,lesieur2017constrained}. Such studies are crucial to isolate individual characteristics of data and machine architecture in a controlled environment and explain their effects on the NN training. For example in \cite{alemanno2023hopfield} the effects of the inference temperature in relation with the dataset's size and noise on RBM learning is investigated. In \cite{theriault2024modelling} the choice of the hidden layer's size in relation to the number of patterns in the data and their correlation is studied. 
In \cite{theriault2024dense,agliari2023denseunsupervised,agliari2023densesupervised}, the role of multi-body interactions is analyzed in the context of Dense Hebbian neural networks.

In this paper, we explore the impact of unit priors on the learning capacity of RBMs and their different learning phases. We extended the results of \cite{alemanno2023hopfield} on the critical dataset's size necessary for an effective learning by characterizing the entire learning phase diagram. We show the existence of a spin-glass phase where the machine learns spourius patterns of information,  an example retrieval phase where the machine tries to learn by memorization and a signal retrieval phase where conversely it learns by generalization. The existence, size and location of these phases strongly depend on the unit priors' choice.  

The paper is organized as follows: Section \ref{sec:1} introduces the teacher-student framework, outlines the main quantities of interest, and describes the class of prior distributions considered.  Section \ref{sec:2} 
presents the main results in terms of a general variational principle for the model's free energy and a system of saddle point equations, whose solutions are connected to our quality of learning estimators. Section \ref{sec:3} presents the Bayesian framework, which is the best possible setting for learning. Finally, Section \ref{sec:4} presents the results in the mismatch architectures between teacher and student RBMs.

\section{The Teacher-Student framework}\label{sec:1}

Consider the inference problem of  a student machine (S-RBM) trained over a dataset generated by a teacher machine (T-RBM). The T-RBM model is built upon $P$ quenched patterns $\boldsymbol{\hat{\xi}}\sim P(\hat{\bxi})$.
A dataset of $M$ examples ${\bm{\mathcal{S}}:=\{\bs^a\}^M_{a=1}}=\{s^a_1,\ldots,s^a_N \}^M_{a=1}$ 
is generated by drawing independent samples from the T-RBM distribution 
\begin{equation}\label{eq:datadistribution}
    P(\bS|\bxip) = \prod_{a=1}^M P(\bs^a|\bxip) = \prod_{a=1}^M z^{-1}(\bxip) P(\bs^a) \ \mathbb{E}_{\hat{\tau}} \exp\left(\sqrt{\frac{\hat{\beta}}{N}}\sum_{i=1}^N\sum_{\mu=1}^{P} s_i^a \hat{\xi}^\mu_i \hat{\tau}_\mu \right)  \;.
\end{equation}
and provided to the student.

\begin{figure}[H]
\centering
\includegraphics[width=.8\textwidth]{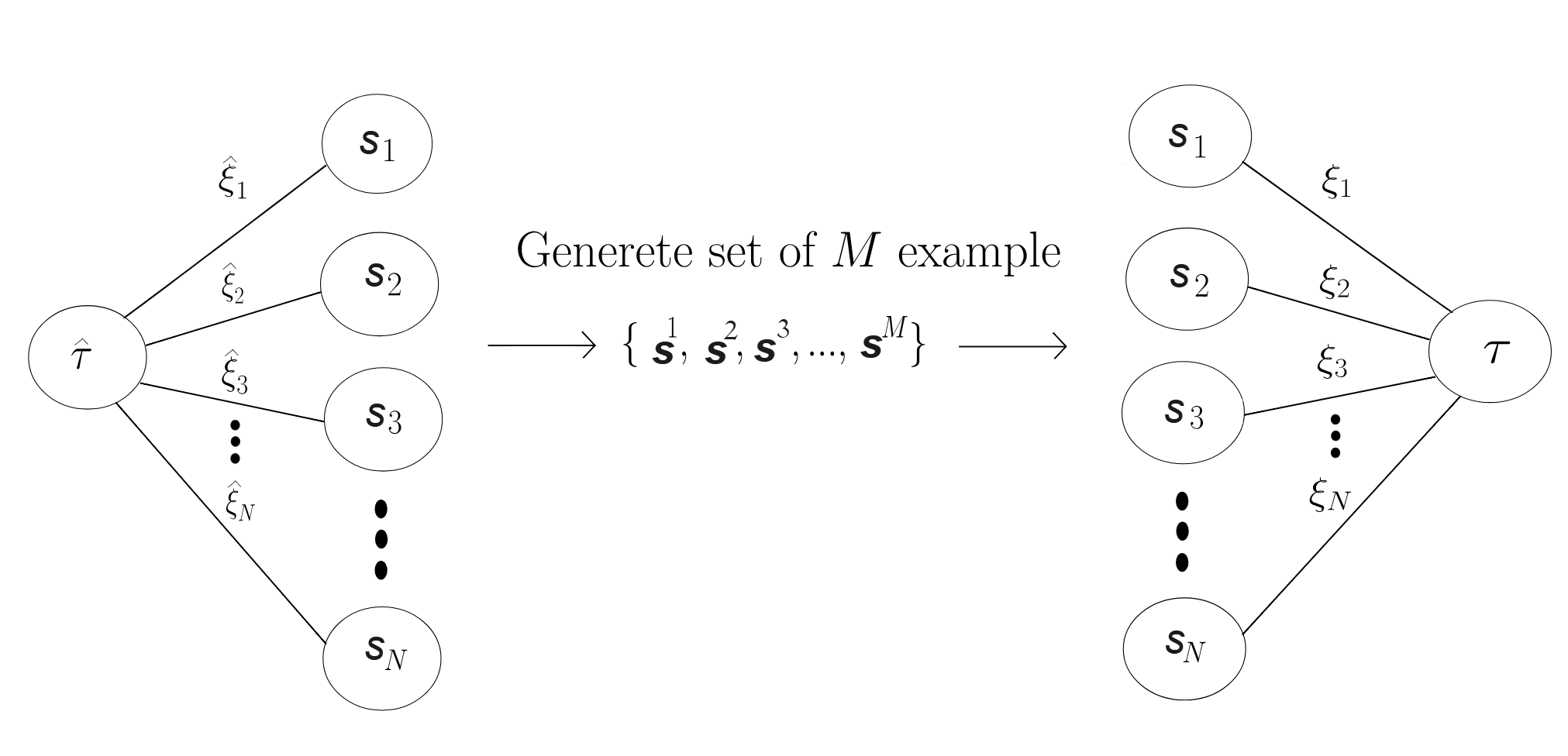} 
    \caption{ Inverse teacher-student problem. On the left, the T-RBM generates the data, following the interactions of the planted signal $\hat{\bm{\xi}}$. On the right, a representation of the S-RBM, which tries to align its own weight vector $\bm{\xi}$ towards $\hat{\bm{\xi}}$ using information extracted from the dataset $\bm{\mathcal{S}}$.}
    \label{fig: TS image}
\end{figure}

The student  is trained over $\bm{\mathcal{S}}$ to recover its structure, i.e. the teacher patterns. In this process the student patterns $\bxi$ are optimized. In a Bayesian framework, they are sampled from the posterior distribution 
\begin{eqnarray}\label{eq:post_prob}
    P(\bm{\xi}|\bm{\mathcal{S}}) &=& \frac{P(\bxi)\prod_{a=1}^M P(\bm{s}^a|\bm{\xi})}{P(\bm{\mathcal{S}})} = Z^{-1}(\bm{\mathcal{S}}) P(\bxi)\prod_{a=1}^M z^{-1}(\bxi)  \ \mathbb{E}_{\tau} \exp\left(\sqrt{\frac{\beta}{N}}\sum_{i=1}^N\sum_{\mu=1}^{P} s_i^a \xi^\mu_i \tau_\mu \right)\nonumber \\
    &=& Z^{-1}(\bm{\mathcal{S}})  z^{-M}(\bxi) \prod_{\mu=1}^P P(\bxi^\mu) \mathbb{E}_{\tau} \exp\left(\sqrt{\frac{\beta}{N}}\sum_{i=1}^N\sum_{a=1}^{M}  \xi^\mu_i s_i^a\tau^a \right)
\end{eqnarray}
with partition function
\begin{equation}
\label{eq: student partition function intro}
    Z(\bm{\mathcal{S}}) = \mathbb{E}_{\bm{\xi}} z^{-M}(\bxi) \prod_{\mu=1}^P\mathbb{E}_{\bm{\tau}} \exp\left( \sqrt{\frac{\beta}{N}} \sum_{i=1}^{N} \sum_{a=1}^{M}\xi^\mu_i {s}_i^a \tau^a  \right)\,.
\end{equation}
In general, the student is unaware of the properties of the T-RBM, therefore we assume $\hat{\beta} \neq \beta$, $P(\bxip)\neq P(\bxi)$, $P(\bm{\hat{\tau}})\neq P(\bm{\tau})$. Without the term $z^{-M}(\bxi)$,  the posterior distribution (\ref{eq:post_prob}) would correspond to $P$ independent generalized Hopfield models, one for each student pattern $\bxi^\mu$, where the examples $\{\bm{s}^a\}_{a=1}^M$ act as dual patterns \cite{barra2017phase, decelle2021inverse, alemanno2023hopfield,theriault2024dense}. In \cite{theriault2024modelling} it is shown that, for teacher patterns that are uncorrelated, the interaction term $z(\bxi)^{-M}$ has the only effect of enforcing orthogonality  between student patterns. Furthermore, the learning performance of a S-RBM with $P$ hidden units learning $P$ teacher patterns is equivalent to that of $P$ separate S-RBM  with a single hidden unit, each learning one pattern. For this reason, in the rest of the paper we assume $P=1$, see Fig. \ref{fig: TS image}.

The \textit{efficiency} of the learning mechanism depends on how effectively the student can recover the ground truth encoded in the teacher's patterns.
The overlap 
\begin{equation}
\label{eqsignal} 
Q(\hat{\bxi},\bxi) = \frac 1 N\sum_{i=1}^N \hat{\xi}_i \xi_i , 
\end{equation}
serves as a reliable measure of learning performance, as it quantifies the similarity between the student’s inferred patterns and the teacher's true patterns. The learning \textit{efficiency} is well known to depend on the data-to-neuron ratio, $\alpha $, the noise level in the examples, $\hat{T} = \hat{\beta}^{-1}$, and the parameter $\beta$, often referred to as the student's inverse inference temperature, which reflects the typical magnitude of the student's weights and acts as a form of regularization.

Our goal is to analyze different scenarios, choosing RBMs with different priors for both teacher and student to observe the effects of these hyper-parameters on the learning efficiency. To this aim, we choose priors from a parametric family of distributions. For each label $x \in \{ \hat{\xi}, \hat{\tau}, \xi, \tau, s\}$, consider the random variable
\begin{equation}
\label{eq:: x interpol}
    \sqrt{\Omega_x} g_x + \sqrt{1-\Omega_x} \epsilon_x \,.
\end{equation}
where $\Omega_x \in [0,1]$,   $g_x \sim N(0,1)$ and $\epsilon_x$ is a Rademacher random variables taking values $\pm 1$. We denote its probability distribution with $P_{\Omega_x}$, which interpolates between a standard gaussian and a binary distributions, having zero mean and unitary variance for any choice of $\Omega_x$. We assume the entries of the teacher pattern $\{\hat{\xi}_i\}_{i=1}^N$ are drawn independently from $P_{\Omega_{\hat{\xi}}}$. Similarly $\hat{\tau}\sim P_{\Omega_{\hat{\tau}}}$,  $\tau\sim P_{\Omega_{\tau}}$,  $\xi_i\sim P_{\Omega_{\xi}}$ and $s^{\mu}_i\sim P_{\Omega_{s}}$. We indicate with $\bm{\Omega}=\{\Omega_x\}_{x\in\{ \hat{\xi}, \tau, \xi, \tau, s\}}$ the set of prior hyperparameters.

We leverage techniques from statistical mechanics to calculate the expected value of the overlap (\ref{eqsignal}) in the limit of large dimension and large dataset $N,M\to\infty$, $M/N=\alpha$ as a function of $\bm{\Omega}$.
Specifically,  we obtain it as a byproduct of the limiting quenched free energy 
\begin{equation}
\label{eq:def_free_entropy}
-\beta f (\bm{\Omega};\alpha, \hat{\beta}, \beta) = \lim_{M,N\to\infty} \frac{1}{N} \mathbb{E}_{\bxip,\boldsymbol{\bm{S}}} \log \left[ Z(\boldsymbol{S}) \right],
\end{equation}
where $\mathbb{E}_{\bxip,\boldsymbol{S}}$ is the expected value w.r.t. the joint distribution of the teacher pattern and generated dataset. In fact, we will show in Section \ref{sec:2} that Eq. (\ref{eq:def_free_entropy}) can be expressed as the result of a variational principle w.r.t. a set of order parameters, whose optimum gives the expected value of the overlap.

\section{Free energy and saddle point equations} \label{sec:2}

The quenched free energy can be computed by exploiting the replica method in the replica symmetry  (RS) approximation and reads as

\begin{equation}
    -\beta f^{RS} (\bm{\Omega}; \alpha,\hat{\beta}, \beta)= -\beta \operatorname{Extr}_{\Lambda, \Lambda_{\tau},\hat{\Lambda},\hat{\Lambda}_{\tau}} \hat{f}(\Lambda, \Lambda_{\tau}, \hat{\Lambda}, \hat{\Lambda}_{\tau}) \,,
\end{equation}
where we introduced the sets of parameters $\Lambda = \{ p, m, q, d \}$ and $\Lambda_{\tau} = \{ p_{\tau}, m_{\tau}, q_{\tau}, d_{\tau} \}$, together with their conjugates $\hat{\Lambda}$ and $\hat{\Lambda}_\tau$. 
The function $\hat{f}$, together with its full derivation, can be found in \ref{Sec:: Appendix A}, specifically in Eq.(\ref{eq:free_energy_RS}). It is extremized by solving the following equations for the order parameters

\begin{gather}\label{eq:saddle_point_equations}
p = \big\langle s \langle \xi \rangle_{\xi|z,s,\hat{\xi}} \big\rangle_{z,s,\hat{\xi}}, \qquad\qquad
p_{\tau} = \big\langle s \langle \tau \rangle_{\tau|z,s,\hat{\tau}} \big\rangle_{z,s,\hat{\tau}} \,,
\\
\label{eq:magn}
m = \big\langle \hat{\xi} \langle \xi \rangle_{\xi|z,s,\hat{\xi}} \big\rangle_{z,s,\hat{\xi}}, \qquad\qquad
m_{\tau} = \big\langle \hat{\tau} \langle \tau \rangle_{\tau|z,s,\hat{\tau}} \big\rangle_{z,s,\hat{\tau}} \,,
\\
\label{eq:overlaps}
q = \big\langle \langle \xi \rangle_{\xi|z,s,\hat{\xi}} ^2\big\rangle_{z,s,\hat{\xi}}, \qquad\qquad
q_{\tau} = \big\langle \langle \tau\rangle_{\tau|z,s,\hat{\tau}} ^2\big\rangle_{z,s,\hat{\tau}}\,,
\\
\label{eq:self_overlaps}
d = \big\langle \langle \xi^2 \rangle_{\xi|z,s,\hat{\xi}} \big\rangle_{z,s,\hat{\xi}}, \qquad
d_{\tau} = \big\langle \langle \tau^2 \rangle_{\tau|z,s,\hat{\tau}} \big\rangle_{z,s,\hat{\tau}} \,.
\end{gather}
The averages $\meanv{.}_{\hat{\xi}}$ and $\meanv{.}_s$ are respectively w.r.t. $P_{\Omega_{\hat{\xi}}}(\hat{\xi})$ and $P_{\Omega_{s}}(s)$, while $\meanv{.}_{\hat{\tau}}$  is over $P_{\Omega_{\hat{\tau}}}(\hat{\tau}) \EXP{\frac{\hat{\beta}}{2}\hat{\tau}^2}$. Finally $\meanv{.}_{\xi|z,s,\hat{\xi}}$ and $\meanv{.}_{\tau|z,s,\hat{\tau}}$ stand for the expectation over respectively 
\begin{equation}
    \label{eq:mean_field_distribution1}
    P_{\Omega_{\xi}}(\xi) \exp\left({ -\frac{\alpha}{2}\beta \left( q_{\tau} - d_{\tau} +\langle \tau^2 \rangle_\tau \right)\xi^{2} + \xi\left(\alpha\sqrt{\hat{\beta}\beta} m_{\tau}\hat{\xi} + \sqrt{\alpha\beta q_{\tau}}z + \beta \Omega_{\tau}p s\right)}\right) \,,
\end{equation}
\begin{equation}
    \label{eq:mean_field_distribution2}
    P_{\Omega_{\tau}}(\tau) \exp{ \left(  -\frac{\beta}{2}\left(q - d\right)\tau^{2} + \tau\left(\sqrt{\hat{\beta}\beta} m \hat{\tau} + \sqrt{\beta q}z + \beta \alpha \Omega p_{\tau}s\right)\right) } \,.
\end{equation}
The optimal order parameters $p,m,q,d$  solving Eqs. (\ref{eq:saddle_point_equations}-\ref{eq:self_overlaps}) have to be interpreted as expected overlaps. First of all, $m$ is the limiting expected overlap between the teacher pattern $\bm{\hat{\xi}}$ and the student pattern $\bm\xi$, i.e. 
 \begin{equation}\label{eq:m}
 m = \lim_{N,M \rightarrow \infty} \mathbb{E}_{\bxip,\boldsymbol{S},\bxi} \left[Q(\bxip,\bxi)\right]
 \end{equation}
where $\mathbb{E}_{\bxip,\boldsymbol{S},\bxi}$ is the expectation w.r.t. the joint distribution $P(\bxip)P(\boldsymbol{S}|\bxip)P(\bxi|\boldsymbol{S})$ of teacher patterns, dataset and student patterns.
Similarly $p$ is the limiting expected overlap between the student pattern $\bm\xi$ and an example $\bm{s}^a$, i.e. 
 \begin{equation}
  p = \lim_{N,M \rightarrow \infty} \mathbb{E}_{\bxip,\boldsymbol{S},\bxi} \left[Q(\bxi,\bm{s}^a)\right].
 \end{equation}
Then $q$ is the limiting expected overlap between any two student patterns $\bxi^{1}$ and $\bxi^{2}$ from two independent posterior samples, i.e. 
\begin{align}\label{eq:q}
q &= \lim_{N,M \rightarrow \infty} \mathbb{E}_{\bxip,\boldsymbol{S}, \bxi^1\times\bxi^2} \left[Q(\bxi^{1},\bxi^{2})\right] 
= \lim_{N,M \rightarrow \infty} \mathbb{E}_{\bxip,\boldsymbol{S}} \left[Q(\mathbb{E}_{\bxi|\boldsymbol{S}}\left[\bxi\right],\mathbb{E}_{\bxi|\boldsymbol{S}}\left[\bxi\right])\right] 
\end{align}
Finally $d$ is the limiting expected self-overlap  i.e. 
\begin{equation}
  d = \lim_{N,M \rightarrow \infty} \mathbb{E}_{\bxip,\boldsymbol{S},\bxi} \left[Q(\bxi,\bxi)\right]\,. 
 \end{equation}
 The RS saddle-point Eqs.(\ref{eq:saddle_point_equations}-\ref{eq:self_overlaps}) can be solved by numerical iteration for any values of the hyper-parameters $\hat{\beta}$, $\beta$, $\alpha$ and $\bm{\Omega}$. We expect the RS solution to be exact when the student is fully informed about the teacher's prior and hyperparameters and matches them with its own, i.e. $\beta = \hat{\beta}$,  $\Omega_{\hat{\xi}}=\Omega_\xi$ and $\Omega_{\hat{\tau}}=\Omega_\tau$. This regime of complete information is commonly referred to as the Bayes Optimal setting or the Nishimori line \cite{nishimori1980exact, iba1999nishimori, nishimori2001statistical, contucci2009spin}. 
 
 Outside of this regime, i.e. in the mismatched setting, replica symmetry breaking corrections are expected at low temperature.
 It is important to remark on the role of $z(\bm \xi)^{-M}$ in Eq. (\ref{eq: student partition function intro}). This term is responsible for the emergence of the quadratic part $-\frac{\alpha \beta}{2}\xi^2 \langle \tau^2 \rangle_{\tau} $ within the distribution (\ref{eq:mean_field_distribution1}). This mechanism enables the automatic regularization of the self-overlap, consistently yielding $d=1$ as the solution to Eq. (\ref{eq:self_overlaps}), as will be shown throughout the following sections, and in \ref{Sec:: Appendix A}.

\section{Exploring arbitrary priors: the Bayesian Optimal setting}\label{sec:3}
With the set of equations (\ref{eq:saddle_point_equations}-\ref{eq:self_overlaps}), we can gain deeper insights into what is happening in the student learning process. In this Section we analyze the ideal situation referred to as the Bayesian optimality setting \cite{barbier2022bayes}, in which the student has access to all the information about the generating process and therefore it is able to mimic the teacher architecture and hyper-parameters to have a better chance of recovering the ground truth $\bm{\hat{\xi}}$. In this regime we therefore assume $\hat{\beta} = \beta$, $\Omega_{\hat{\xi}}=\Omega_\xi$ and $\Omega_{\hat{\tau}}=\Omega_\tau$. At low temperature all the examples have macroscopic alignment with the signal and the student can easily learn it perfectly. Conversely, at a sufficiently high generating temperature $\hat{\beta}$ the examples have vanishing overlap with the teacher pattern. At the same time, the high inference temperature $\beta=\hat{\beta}$ prevents the student pattern to have macroscopic alignment with a single example, i.e. $p=0$. The saddle point equations (\ref{eq:saddle_point_equations}-\ref{eq:self_overlaps}) reduces to
\begin{gather}
\label{eq:: Bayes optimal MF equations}
m = \big\langle \hat\xi \langle \xi   \rangle_{\xi|z,s,\hat\xi} \big\rangle_{z,s,\hat\xi}
\qquad \qquad
m_{\tau} = \big\langle \hat\tau \langle \tau   \rangle_{\tau|z,s,\hat\tau} \big\rangle_{z,s,\hat\tau}
\\
q = \big\langle \langle \xi \rangle_{\xi|z,s,\hat\xi}^2 \big\rangle_{z,s,\hat\xi}
\qquad\qquad
q_{\tau} = \big\langle \langle \tau  \rangle_{\tau|z,s,\hat\tau}^2 \big\rangle_{z,s,\hat\tau}
\\
d = \big\langle \langle \xi^2  \rangle_{\xi|z,s,\hat\xi} \big\rangle_{z,s,\hat\xi}
\qquad\qquad
d_{\tau} = \big\langle \langle \tau^2   \rangle_{\tau|z,s,\hat\tau} \big\rangle_{z,s,\hat\tau} \;.
\end{gather}
We can further reduce the number of equations thanks to the Nishimori identities \cite{contucci2009spin,iba1999nishimori,nishimori1980exact}. Infact, in the Bayes optimal setting it holds
\begin{equation}\label{eq:nishi}
\mathbb{E}_{\bxip,\bm{S},\bxi^1\times\ldots\times\bxi^k}  f(\bxi^1,\ldots,\bxi^k)
    =\,
    \mathbb{E}_{\bxip,\bm{S},\bxi^2\times\ldots\times\bxi^k}  f(\bxip,\bxi^2,\ldots,\bxi^k),
\end{equation}
for any regular function $f$ and with $(\bxi^1,\ldots,\bxi^k)$ being $k$ independent samples from the posterior distribution. In particular we can apply the identity (\ref{eq:nishi}) to the overlap obtaining that 
\begin{equation}\label{eq:nishi1}
\mathbb{E}_{\bxip,\bm{S},\bxi^1\times\bxi^2}  Q(\bxi^1,\bxi^2)
    =\,
    \mathbb{E}_{\bxip,\bm{S},\bxi}  Q(\bxip,\bxi).
\end{equation}
As a consequence, recalling Eqs. (\ref{eq:m},\ref{eq:q}),  it must be $m=q$ and
therefore it is sufficient to solve 
\begin{gather}
\label{eq: m BO}
m = \big\langle \xi \langle \xi   \rangle_{\xi|z,s} \big\rangle_{z,s}
\qquad \qquad
m_{\tau} = \big\langle \tau \langle \tau   \rangle_{\tau|z,s} \big\rangle_{z,s} \\
\label{eq: d BO}
d = \big\langle \langle \xi^2  \rangle_{\xi|z,s} \big\rangle_{z,s}
\qquad\qquad
d_{\tau} = \big\langle \langle \tau^2   \rangle_{\tau|z,s} \big\rangle_{z,s} \;,\end{gather}
where the averages are now recast in the form
\begin{equation}
    \label{eq:mean_field_distribution1 BO}
    P_{\Omega_{\xi}}(\xi) \exp{ \left( -\frac{\alpha}{2}\beta \left(m_{\tau} - d_{\tau} +\langle \tau^2 \rangle_\tau \right)\xi^{2} + \xi  \sqrt{\alpha\beta m_{\tau}}z \right) } \,,
\end{equation}
\begin{equation}
\label{eq:mean_field_distribution2 BO}
    P_{\Omega_{\tau}}(\tau) \exp{ \left( -\frac{\beta}{2}\left(m - d\right)\tau^{2} + \tau \sqrt{\beta m}z \right)} \,.
\end{equation}
\begin{figure}[H]
\centering
\begin{subfigure}{.6\textwidth}
  \centering
\hspace{-3.0cm}\includegraphics[width=.9\linewidth]{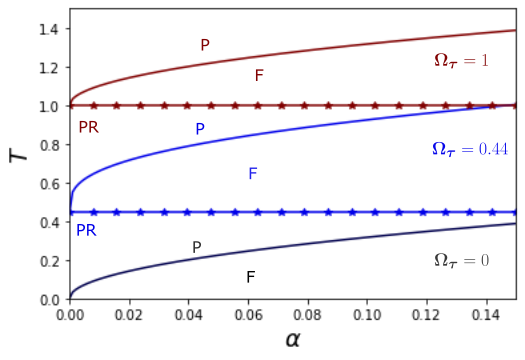}
\end{subfigure}%
\begin{subfigure}{.6\textwidth}
  \centering
\hspace{-3.0cm}\includegraphics[width=.9\linewidth]{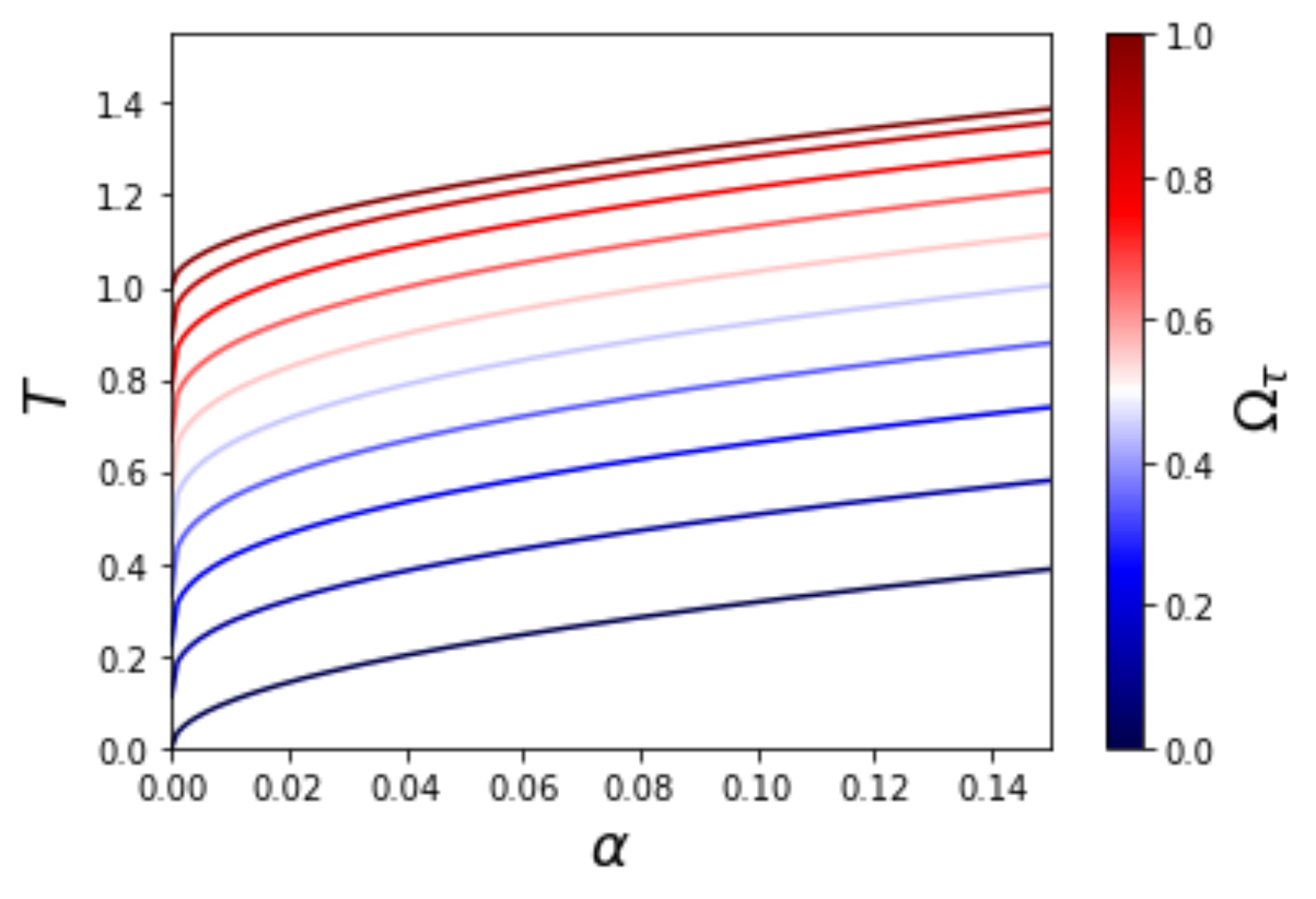}
\end{subfigure}%
  \caption{ Retrieval phase transition lines in the case of the Bayesian optimal scenario. Below the transition curve it is possible to recover the planted signal $\bm{\hat{\xi}}$. \textbf{Left}: P-F transition line for $\Omega_{\tau}\in\{0,0.44,1\}$. For each choice of the $\tau$ prior the perfect retrieval (PF) region starts  below the star marked line, i.e. $T<\Omega_\tau$. \textbf{Right}: Different P-F lines (without the perfect retrieval region) for all the possible values of $\Omega_\tau$. Above each colored line (taken singularly) the student is in a paramagnetic phase, below in a ferromagnetic regime. 
}\label{fig: BayesOptimal}
\end{figure}
 In Fig. (\ref{fig: BayesOptimal})  the learning phase diagram in the $(\alpha,T)$ plane is presented as a function of $\bm{\Omega}$. One can observe three different regimes: 
\begin{itemize}
\item the \textit{Paramagnetic}  (P) phase, where the order parameter $m=q$ vanishes;
\item the \textit{ Ferromagnetic} (F) phase, where $m>0$.  
\item  the \textit{ Perfect Retrieval} (PR) phase, where $m=1$. 
\end{itemize}
In the ferromagnetic phase the student RBM is capable of learning (at least partially) the teacher pattern. Conversely this ability is lost in the paramagnetic phase. The ferromagnetic phase emerges even at high temperature once the dataset size reaches a critical threshold, i.e.  $\alpha\geq\alpha_c (T,\bm{\Omega})$. In fact, at high generating temperature, each example provides only a vanishing amount of information about the teacher pattern. Therefore an extensively large number of examples are needed for the student to retrieve the signal. The corresponding P-F phase transition is thus the onset of learning. 
Descending from high temperatures, where $m=m_\tau = 0$, the P-F transition line can be obtained by expanding in  $m\sim0$. First of all, note that at $m=m_\tau = 0$ Eqs. (\ref{eq: m BO}, \ref{eq: d BO}) give
\begin{eqnarray}
    && d_{\tau}= \langle \tau^2 \rangle_0  =  \mathbb{E}_\tau[ \tau^2 \exp{\left(\frac{\beta}{2}   d \,\tau^{2} \right)}] / \mathbb{E}_\tau[ \exp{ \left(\frac{\beta}{2} d \,\tau^{2} \right)  }] \,,  \qquad  \tau \sim P_{\Omega_\tau}\label{eq: dt BO}\\
    && \hspace{-2.2cm}d = \langle \xi^2 \rangle_0 =  \mathbb{E}_\xi[ \xi^2 \exp{ \left( \frac{\alpha \beta}{2}  (d_\tau -\langle \tau^2\rangle_\tau)  \,\xi^{2} \right) }] / \mathbb{E}_\xi[ \exp{ \left( \frac{\alpha \beta}{2}(d_\tau -\langle \tau^2\rangle_\tau) \,\xi^{2} \right)  }]\,, \ \   \xi \sim P_{\Omega_\xi}
\end{eqnarray}
where $\langle \cdot \rangle_{0}$  denotes the expectations w.r.t. the distributions of Eqs. (\ref{eq:mean_field_distribution1 BO}) or (\ref{eq:mean_field_distribution2 BO}) at zero effective field. Note that, since $d_{\tau} = \langle \tau^2 \rangle_\tau$ it is $ \EXP{\frac{\alpha \beta}{2}(d_\tau - \langle \tau^2 \rangle_\tau) \,\xi^2} = 1$ and therefore $d=1$. This is expected because, thanks to the Nishimori identities, it must be 
\begin{equation}
\mathbb{E}_{\bxip,\bm{S},\bxi}  f(\bxi)
    =\,
    \mathbb{E}_{\bxip,\bm{S},\bxi}  f(\bxip)= \mathbb{E}_{\bxip}  f(\bxip).
\end{equation}
Therefore the statistics of the student pattern matches those of the teacher's: in particular the self-overlap must be equal to that of the teacher pattern, being one by definition.
By expanding the effective distributions in Eqs. (\ref{eq:mean_field_distribution1 BO}, \ref{eq:mean_field_distribution2 BO}) as
\begin{eqnarray}
&& P_{\Omega_{\xi}}(\xi) \left(1+ \sqrt{\alpha\beta m_\tau} z \, \xi +O(q_\tau)\right) \\
&& P_{\Omega_{\tau}}(\tau) \left(1+ \sqrt{\beta m} z \, \tau +O(q)\right) \exp{\left\{\frac{\beta}{2}\tau^2 \right\} } \,
\end{eqnarray}
one obtains
\begin{eqnarray}
m &=& \big\langle \xi \langle \xi \rangle_{\xi|z,s} \big\rangle_{z,s} = \alpha \beta m_\tau +o(m_\tau)\\
m_{\tau} &=& \big\langle \tau\langle \tau \rangle_{\tau|z,s}\big\rangle_{z,s}= \beta m \langle \tau^2 \rangle_{0}^2 +o(m) \,.
\end{eqnarray}
As a consequence the transition line turns out to be 
\begin{equation}
\label{eq: transition BO}
    1 = \alpha \beta^2  d_{\tau}^2 \,.
\end{equation}
Eqs. (\ref{eq: dt BO}) and (\ref{eq: transition BO}) can be solved numerically obtaining all the P-F transition curves in Fig.(\ref{fig: BayesOptimal}). Note that the transition $T_c(\bm\Omega,\alpha)$ is only function of $\Omega_{\tau}$ and that it is an increasing function of $\Omega_{\tau}$ and $\alpha$. In particular, in the low load limit, $\alpha{\to 0}$, the transition temperature is $T_c(\bm{\Omega},0)=\Omega_{\tau}$.
At lower temperatures, specifically when $T \leq T_c(\bm{\Omega},0)$ the Perfect Retrieval regime appears. In this region,  since each example exhibits a finite overlap with the teacher pattern, the student can perfectly retrieve the signal, i.e.  $m = 1$,  for any extensive dataset $\alpha > 0$. 

\section{Exploring arbitrary priors: Mismatched Setting}\label{sec:4}
In a more realistic scenario, the student is not aware of the model underlying the data. In a teacher student framework it means that the inference temperature is in general different from the dataset noise, i.e. $\beta\neq\hat{\beta}$ and that the hyperparameters of the S-RBM do not match the ones of the T-RBM, i.e. $\Omega_\xi\neq\Omega_{\hat{\xi}}$ and $\Omega_\tau\neq\Omega_{\hat{\tau}}$. 
In this situation the phase diagram is characterized by four different regimes:
\begin{itemize}
    \item the \textit{Paramagnetic} (P) region, where $m=q=p=0$;
    \item the \textit{Signal Retrieval} (sR) region, where $m\neq 0$, $q>0$, $p=0$;
    \item the \textit{Example retrieval} (eR) region, where $p\neq 0$, $q>0$, $m=0$;
    \item the \textit{Spin Glass} (SG) region, where $m=p=0$, $q>0$.
\end{itemize}
In the eR phase the student is not capable of learning the teacher pattern because it is aligned with one example that shares vanishing overlap with the signal.
In the SG phase, the student retrieves a spurious pattern unrelated to both the teacher and any examples. The sR phase is the only region where learning is possible, with phase transitions marking the boundaries between different learning regimes that depend on $\bm{\Omega}$.

\subsection{Transition to the Spin Glass Phase} 
At very high temperature ($\beta \sim 0$) the distributions (\ref{eq:mean_field_distribution1}) and (\ref{eq:mean_field_distribution2}) have no external effective fields and therefore $\meanv{\xi}=\meanv{\tau}=0$. In this regime the student is just making a random guess. All the relevant order parameters $p=m=q =0$ together with their hidden counterparts $p_{\tau}= m_{\tau}= q_{\tau} =0$,  i.e. the system is in the paramagnetic phase.
As in the Bayesian Optimal setting, lowering the temperature may start a spin glass transition, with nonzero
overlaps $q$ and $q_\tau$; while the other parameterns remain zero. Assuming this transition is continuous, we can linearise
Eqs. (\ref{eq:overlaps}) for small $q$.
Expanding first Eqs. (\ref{eq:mean_field_distribution1}) and (\ref{eq:mean_field_distribution2}) we get that $\xi$ and $\tau$ in Eqs. (\ref{eq:overlaps}) have to be averaged out over respectively
\begin{eqnarray}
&& P_{\Omega_{\xi}}(\xi) \left(1+ \sqrt{\alpha\beta q_\tau} z \xi +O(q_\tau)\right) \EXP{\frac{\alpha}{2}\beta  (d_{\tau} - \langle \tau^2 \rangle_\tau) \xi^{2} } \\
&& P_{\Omega_{\tau}}(\tau) \left(1+ \sqrt{\beta q} z \tau +O(q)\right) \EXP{\frac{\beta}{2} d\,\tau^2} \,.
\end{eqnarray}
First of all it is 
\begin{eqnarray}
d_\tau=\langle \tau^2 \rangle_{0} =  \mathbb{E}_\tau[ \tau^2 \EXP{\frac{\beta}{2}  d \, \tau^{2} }] / \mathbb{E}_\tau[ \EXP{\frac{\beta}{2}d \,\tau^{2}}],\ \ \ \ \ \ \tau \sim P_{\Omega_\tau}
\label{eq:selfoverl_tr1}
\end{eqnarray}
\begin{equation}
    \label{eq:selfoverl_tr}\hspace{-1.5cm}d=\langle \xi^2 \rangle_{0} =  \mathbb{E}_\xi [\xi^2 \EXP{\frac{\alpha \beta}{2}  (d_{\tau} -\langle \tau^2 \rangle_\tau) \xi^{2} }] / \mathbb{E}_\xi[\EXP{\frac{\alpha}{2}\beta  (d_{\tau} - \langle \tau^2 \rangle_\tau)\xi^{2} }],\ \ \ \xi\sim P_{\Omega_\xi}\,,
\end{equation}
where $\langle \cdot \rangle_{0}$  denotes the expectations w.r.t. the distributions of Eqs. (\ref{eq:mean_field_distribution1}) or (\ref{eq:mean_field_distribution2}) at zero effective field (i.e. $m=m_\tau=q=q_\tau=p=p_\tau=0$). Therefore also in this case it is $d=1$ at the transition. Moreover we get 
\begin{eqnarray}
q &=& \big\langle \langle \xi \rangle_{\xi|z,s,\hat{\xi}} ^2\big\rangle_{z,s,\hat{\xi}} = \alpha \beta q_\tau  \langle \xi^2 \rangle_{0} ^2 + o(q_\tau)= \alpha \beta q_\tau + o(q_\tau)  \label{eq:qlin1}\\
q_{\tau} &=& \big\langle \langle \tau \rangle_{\tau|z,s,\hat{\tau}}^2 \big\rangle_{z,s,\hat{\tau}}= \beta q  \langle \tau^2 \rangle_{0}^2 +o(q)=\beta q d_\tau^2 +o(q),\label{eq:qlin2}
\end{eqnarray}
From Eqs. (\ref{eq:qlin1}-\ref{eq:qlin2}) a second order phase transition to the spin glass phase may occur at 
\begin{equation} \label{eq:: SG transition} 1 = \alpha \beta^2 d_{\tau}^2 \;
\end{equation} 
that can be solved to obtain the critical temperature  $T_c(\alpha)$ or equivalently the critical size $\alpha_c(T)$. Note that there is no dependence of $\hat{\beta}$, meaning that the spin glass transition only depends on a (too low) inference temperature of the student. As expected \cite{barra2017phase,alemanno2023hopfield,theriault2024dense} 

the P-SG transition (\ref{eq:: SG transition}) corresponds to the P-F transition in the Nishimori regime (\ref{eq: transition BO}). Moreover it also corresponds to the P-SG transition of a generalized Hopfield model with random patterns at inverse temperature $\beta$ and load $\alpha$. Indeed, the expression (\ref{eq:: SG transition}) matches the one found in \cite{barra2018phase}.

For example fixing $\Omega_{\xi}=\Omega_{\tau}=0$ one gets the bipartite SK P-SG transition $T_c(\alpha) = \sqrt{\alpha}$, while using $\Omega_{\xi}=0$ and $\Omega_{\tau}=1$ it is possible to recover the P-SG transition line of the Hopfield model $T_c(\alpha)=1+\sqrt{\alpha}$. 

\subsection{Transition to the Retrieval Phase}
When the size of the dataset is sufficiently large, lowering the temperature may lead to a continuous transition towards a signal retrieval (sR) region. The transition line can be obtained by expanding Eq. (\ref{eq:magn}) for small $m$ and $m_\tau$, keeping $p=p_\tau=q=q_\tau=0$. As previously done, first of all Eqs. (\ref{eq:mean_field_distribution1}-\ref{eq:mean_field_distribution2}) can be expanded as 
\begin{eqnarray}
&& P_{\Omega_{\xi}}(\xi) \left(1+ \alpha\sqrt{\beta \hat{\beta} } m_\tau \hat{\xi}\xi +O(m_\tau)\right) \EXP{\frac{\alpha}{2}\beta  (\,d_{\tau}-\langle \tau^2 \rangle_\tau\,)\xi^{2} } \\
&& P_{\Omega_{\tau}}(\tau) \left(1+ \sqrt{\beta \hat{\beta} } m \tau\hat{\tau} +O(m)\right) \EXP{\frac{\beta}{2} d\tau^2}, 
\end{eqnarray}
from which it holds 
\begin{align}
\hspace{-2cm}
m &= \big\langle \hat{\xi} \langle \xi \rangle_{\xi|z,s,\hat{\xi}} \big\rangle_{z,s,\hat{\xi}} = \alpha \sqrt{\hat{\beta}\beta} m_\tau \langle\hat{\xi}^2\rangle_{\hat{\xi}} \langle \xi^2 \rangle_{0}  + \mathcal{O}(m_\tau)=\alpha \sqrt{\hat{\beta}\beta} \, m_\tau \,d \,\langle \hat\xi\rangle_{\hat\xi}  + \mathcal{O}(m_\tau)  \label{eq:mlin1}\\
m_{\tau} &= \big\langle \hat{\tau}\langle \tau \rangle_{\tau|z,s,\hat{\tau}}\big\rangle_{z,s,\hat{\tau}}= \sqrt{\hat{\beta}\beta} m\langle\hat{\tau}^2\rangle_{\hat{\tau}} \langle \tau^2 \rangle_{0} +\mathcal{O}(m)=\sqrt{\hat{\beta}\beta} \,m \,d_\tau \langle\hat{\tau}^2\rangle_{\hat{\tau}} +\mathcal{O}(m),\label{eq:mlin2}
\end{align}
where again $\langle \cdot \rangle_{0}$ denotes the expectations w.r.t. the distributions of Eqs. (\ref{eq:mean_field_distribution1}) or (\ref{eq:mean_field_distribution2}) at zero effective field. In Eq.(\ref{eq:mlin1}) we can use  $ d_{\hat{\xi}}:=\langle\hat{\xi}^2\rangle_{\hat{\xi}}  =1$ which derives from the analysis of the teacher partition function (\ref{Aeq:: n^0 order params}). Again the self-overlaps $d$ and $d_\tau$ are solutions of Eq.(\ref{eq:selfoverl_tr1}-\ref{eq:selfoverl_tr}), which implies $d=1$. Recalling that $\meanv{.}_{\hat{\tau}}$  is the average over the tilted distribution $P_{\Omega_{\hat{\tau}}}(\hat{\tau}) \EXP{\frac{\hat{\beta}}{2}\hat{\tau}^2}$, it holds
\begin{eqnarray}
  d_{\hat{\tau}}&:=&\langle\hat{\tau}^2\rangle_{\hat{\tau}} =  \frac{1-\hat{\beta}\Omega_{\hat{\tau}}^2}{(1-\hat{\beta}\Omega_{\hat\tau})^2}\,,
\end{eqnarray}
thus, from Eqs. (\ref{eq:mlin1}-\ref{eq:mlin2}), a second order phase transition to the sR phase may occur at 
\begin{equation} \label{eq:: sR transition} 1 = \alpha \hat{\beta}\beta d_{\tau}d_{\hat{\tau}}\,.
\end{equation} 
 For example when both the T-RBM and S-RBM have only binary units the critical temperature reads as  $
T_c(\alpha) = \alpha \hat{\beta}$ while in the case of the classical Hopfield model, i.e. $\Omega_\tau=\Omega_{\hat{\tau}}=1$ \cite{alemanno2023hopfield},
$
T_c(\alpha) = 1+ \alpha \hat{\beta} \;.
$
Since $d=1$ and  $d_{\tau}$ doesn't depend of $\alpha$, the P-sR critical temperature $T_c(\alpha)$  grows linearly with the dataset's size $\alpha$. Conversely, from Eq. (\ref{eq:: SG transition}), the critical temperature of the P-SG transition grows only as $O(\sqrt{\alpha})$. This means that for larger values of $\alpha$ the P-sR phase transition occurs first. On the contrary for small $\alpha$ the  spin glass phase emerges at a higher temperature.

\begin{figure}[H]
\centering
\begin{subfigure}{.7\textwidth}
  \centering
\hspace{-5cm}\includegraphics[width=0.9
\linewidth]{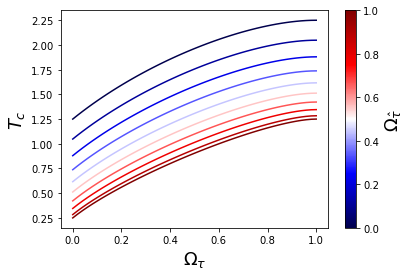}
\end{subfigure}%
\begin{subfigure}{.7\textwidth}
  \centering
\hspace{-6cm}\includegraphics[width=0.9\linewidth]{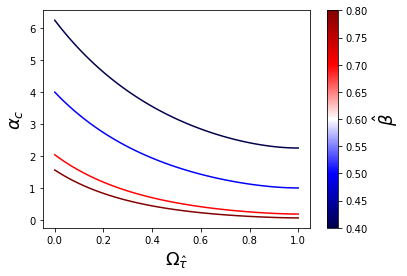}
\end{subfigure}%
 \caption{
 Triple point's coordinates as a function of the prior parameters. \textbf{Left}: Critical temperature for different choices of the the hidden units prior of both S-RBM and T-RBM. $T_c$ is an increasing funciton of $\Omega_\tau$, but decreases with $\Omega_{\hat\tau}$. \textbf{Right}: Critical size as function of the relevant teacher variables. It is a decreasing function of both $\Omega_{\hat\tau}$ and $\hat T=1/\hat\beta$.
 }\label{fig: Tcrit}
\end{figure}

A triple point exists when the two transition lines (\ref{eq:: SG transition}) and (\ref{eq:: sR transition}) cross, i.e. at the solution of 
\begin{equation}
\label{eq: crit system}
\begin{cases}
1= \alpha \beta^2 d_{\tau}^2  \\
1 = \alpha \beta \hat{\beta}d_{\tau}d_{\hat{\tau}} , 
\end{cases}
\end{equation}
giving 
\begin{equation} \label{eq: critical load}
    P_{triple}=(\alpha_c,T_c)=\left(
    \frac{1}{\hat{\beta}^2 d_{\hat{\tau}}^2}\ ,\ 
    \frac{d_{\tau} }{\hat{\beta} d_{\hat{\tau}}}\right).
\end{equation}
As expected the Nishimori line $\beta=\hat{\beta}$ crosses the triple point when the student and teacher priors match, i.e. $d_\tau=d_{\hat{\tau}}$. It is interesting to note that the $\alpha$ of the triple point only depends on the properties of the data, i.e. of the teacher, while its temperature may also depend on the student's priors.
The critical temperature at the triple point, $T_c$, is shown in Fig. \ref{fig: Tcrit} to be an increasing function of $\Omega_{\tau}$. 
Explicitly, $T_c$ increases with $\Omega_{\tau}$, but decreases with $\Omega_{\hat\tau}$. Conversely, the critical value $\alpha_c$ decreases as both $\hat T$ and $\Omega_{\hat\tau}$ increase. 

These properties can be observed from the phase diagrams obtained by different choices of the priors, which are presented in Figs. (\ref{fig: OTvariations}-\ref{fig: Data prior variation}).
All the depicted transition lines from the SG region to the sR or the eR regions can be found numerically by solving the saddle point Eqs. (\ref{eq:saddle_point_equations}-\ref{eq:self_overlaps}).

 In all the cases the triple point appears to be the point of the sR region with smaller $\alpha$. In other words the $\alpha$ of the triple point corresponds to the minimum size of the dataset for which learning is possible, at least for suitable choice of the inference temperature. For this reason we refer to it as the model's  $\alpha_c$. For $\alpha\geq \alpha_c$ there exists an optimal inference temperature where the learning performance, i.e. the overlap with the teacher pattern, is maximal. Notably if the inference temperature is set too low one can eventually exit the sR region and enters a glassy phase. However, for sufficiently large values of $\alpha$, the sR region extends down to zero temperature. 

Finally, the example retrieval region may emerge at low inference temperature when $\alpha$ is very close to zero. This phase corresponds to the pattern retrieval region in the generalized Hopfield models \cite{barra2018phase} under the critical capacity. In this region the student pattern tends directly to be aligned with one of the examples instead of learning their archetype, i.e. the teacher pattern. Unlike the sR regime, this is a phase where the machine is trying to learn  by memorization and not by generalization \cite{alemanno2023hopfield}.

\subsection{The role of the S-RBM priors}
By fixing the properties of the T-RBM, i.e. of the generating process, it is possible to explore the effects of the S-RBM priors on the learning performance. In the following discussion the examples are assumed to have binary entries, i.e. $\Omega_s=0$. In the following section, as well as in the subsequent sections, the phase diagrams presented illustrate the existing phases.
 
\begin{figure}[H]
\centering
\begin{subfigure}{.6\textwidth}
  \centering
\hspace{-8.0cm}\includegraphics[width=.75\linewidth]{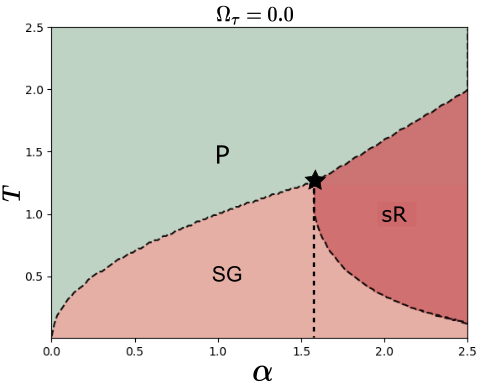}
\end{subfigure}%
\begin{subfigure}{.6\textwidth}
  \centering
\hspace{-12.0cm}
\vspace{-0.2cm}\includegraphics[width=0.8\linewidth]{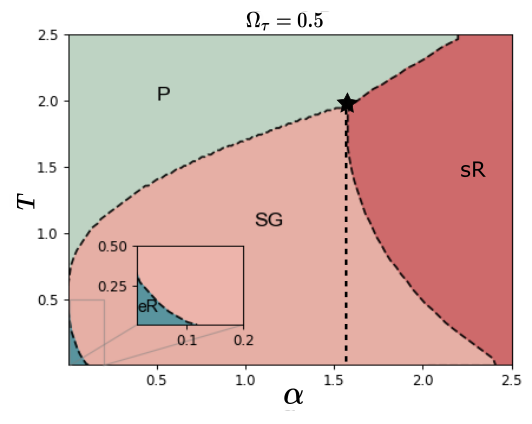}
\end{subfigure}%
\begin{subfigure}{.6\textwidth}
  \centering
\hspace{-16.0cm}\includegraphics[width=.75\linewidth]{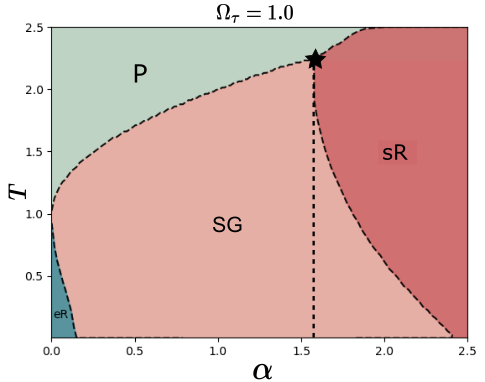}
\end{subfigure}
  \caption{Phase diagram of different S-RBM configurations, each with a different choice of hidden unit prior $\Omega_{\tau}$. All Students have a binary prior for $\xi$, i.e., $\Omega_\xi=0$. The teacher is generating ($\hat\beta=0.8$) binary data ($\Omega_s=0$), and its architecture is fixed by the choice $\Omega_{\hat\xi}=\Omega_{\hat\tau}=0$. The black star represents the critical point $(\alpha_c, T_c)$, while the vertical dashed line shows that the position $\alpha_c \simeq 1.55$ is the same for all the images. The eR phase emerges when the $\tau$ prior has a gaussian tail.} \label{fig: OTvariations}
\end{figure}
In Fig. (\ref{fig: OTvariations}) the phase diagrams of different S-RBMs with binary patterns ($\Omega_\xi=0$) are shown in the case of a binary T-RBM, i.e. $\Omega_{\hat{\xi}}=\Omega_{\hat{\tau}}=0$. One can observe the appearance of the eR phase at low temperature as soon as the student hidden unit starts to have a Gaussian tail, i.e. $\Omega_\tau>0$. The critical temperature at $\alpha=0$ is exactly $\Omega_\tau$. In the limit $\Omega_\tau\to 1$ this phase extends up to the critical capacity of the Hopfield model $\alpha=0.14$. As expected the critical size $\alpha_c$ of the triple point doesn't change with the student priors while its temperature increases with $\Omega_\tau$. As a consequence the area of the \textit{sR} phase expands in the direction of higher temperatures. This is typically an advantage 
 for MC algorithms that samples from the posterior starting from high temperature like simulated annealing \cite{baldassi2018efficiency,hinton2012practical}. 
 
\begin{figure}[H]
\centering
\begin{subfigure}{.6\textwidth}
  \centering
\hspace{-8.0cm}\includegraphics[width=.75\linewidth]{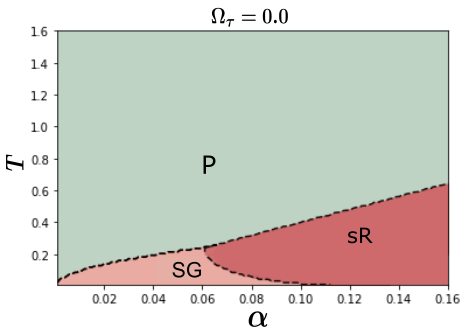}
\end{subfigure}%
\begin{subfigure}{.6\textwidth}
  \centering
\hspace{-12.0cm}\includegraphics[width=.75\linewidth]{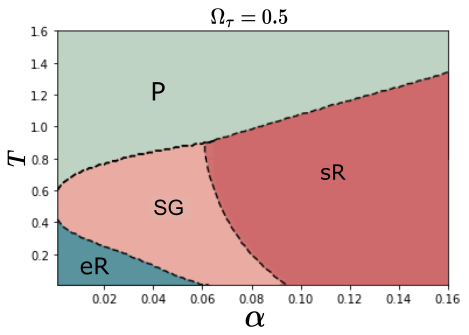}
\end{subfigure}%
\begin{subfigure}{.6\textwidth}
  \centering
\hspace{-16.0cm}\includegraphics[width=.75\linewidth]{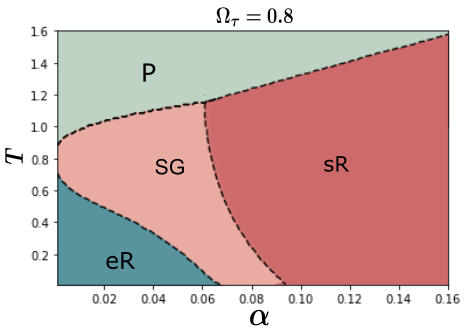}
\end{subfigure}
  \caption{ Phase diagrams showing the effect of student's hidden unit prior when the teacher generates ($\hat\beta=0.8$) the dataset ($\Omega_s = 0$) using a Hopfield model. The student pattern is chosen with $\Omega_\xi=0$. The critical size needed to enter the inference regime is unaltered by the choice of $\Omega_\tau$ and its value is $\alpha_c \simeq 0.06$, while the temperature $T_c$ increases.}\label{fig: hopfieldTeacher OTvariations}
\end{figure}
In Fig. (\ref{fig: hopfieldTeacher OTvariations}), we observe similar behavior for a dataset generated by a Hopfield like T-RBM, i.e. $\Omega_{\hat{\tau}}=1$, $\Omega_{\hat{\xi}}=0$. The different generation procedures significantly affect the number of data points required for the student to transition into the sR regime: as prescribed by Eq.(\ref{eq: critical load}), $\alpha_c$ is smaller than that of  Fig.(\ref{fig: OTvariations}). One can in principle investigate the effect of changing $\Omega_\xi$, however we have verified that the phase diagram is not affected by that prior since it only enters in the self-overlap,  and  $d=1$ is always solution of Eqs. (\ref{eq:saddle_point_equations}-\ref{eq:self_overlaps}). Interestingly, the benefit of using Gaussian units—where at any given temperature, the sR region occurs at smaller $\alpha$—remains consistent regardless of the dataset properties. Specifically, it is not always advantageous for the student to align its hyperparameters with those of the teacher.

\begin{figure}[H]
\centering
\begin{subfigure}{.6\textwidth}
  \centering
\hspace{-8.0cm}\includegraphics[width=.75\linewidth]{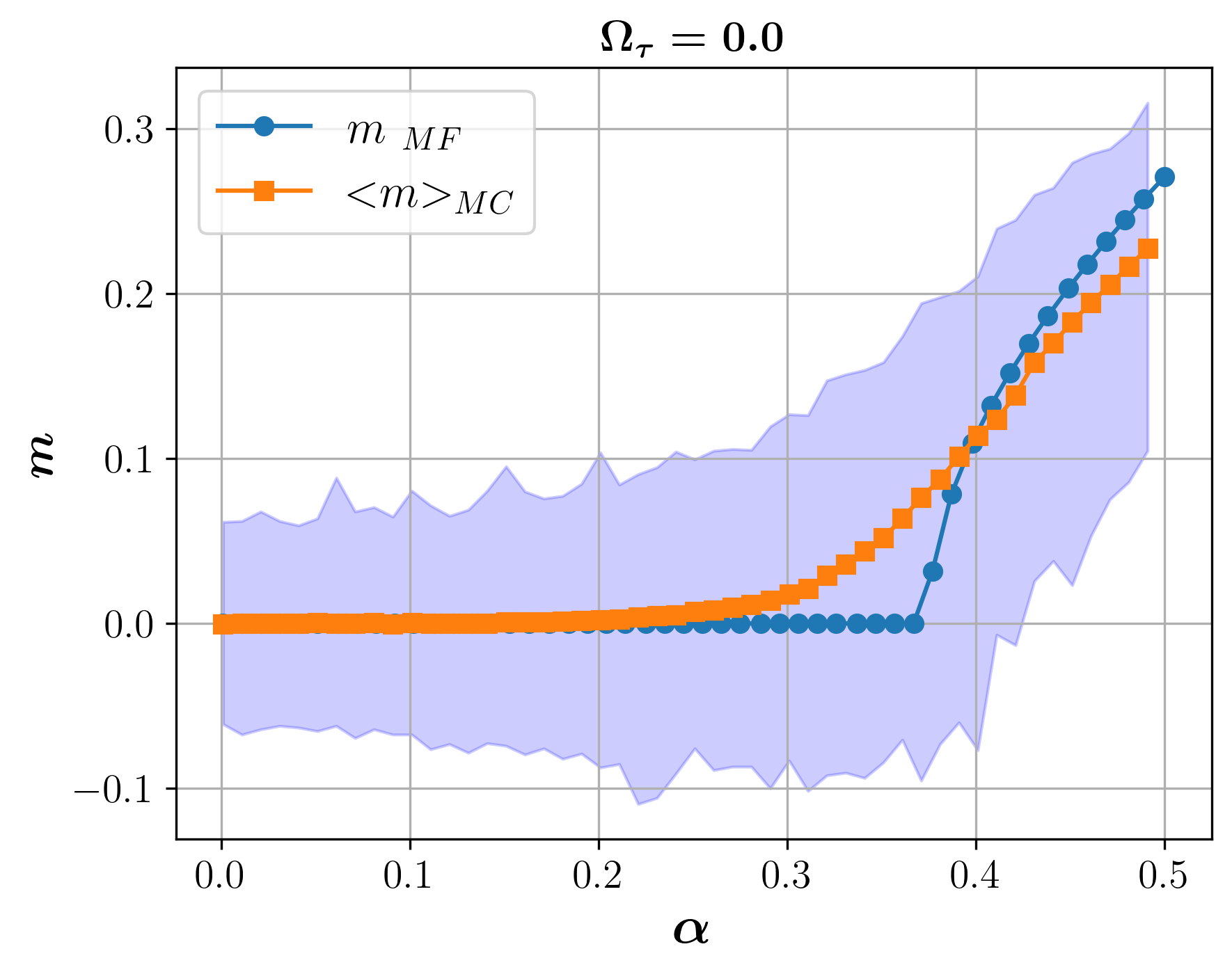}
\end{subfigure}%
\begin{subfigure}{.6\textwidth}
  \centering
\hspace{-12.0cm}\includegraphics[width=.75\linewidth]{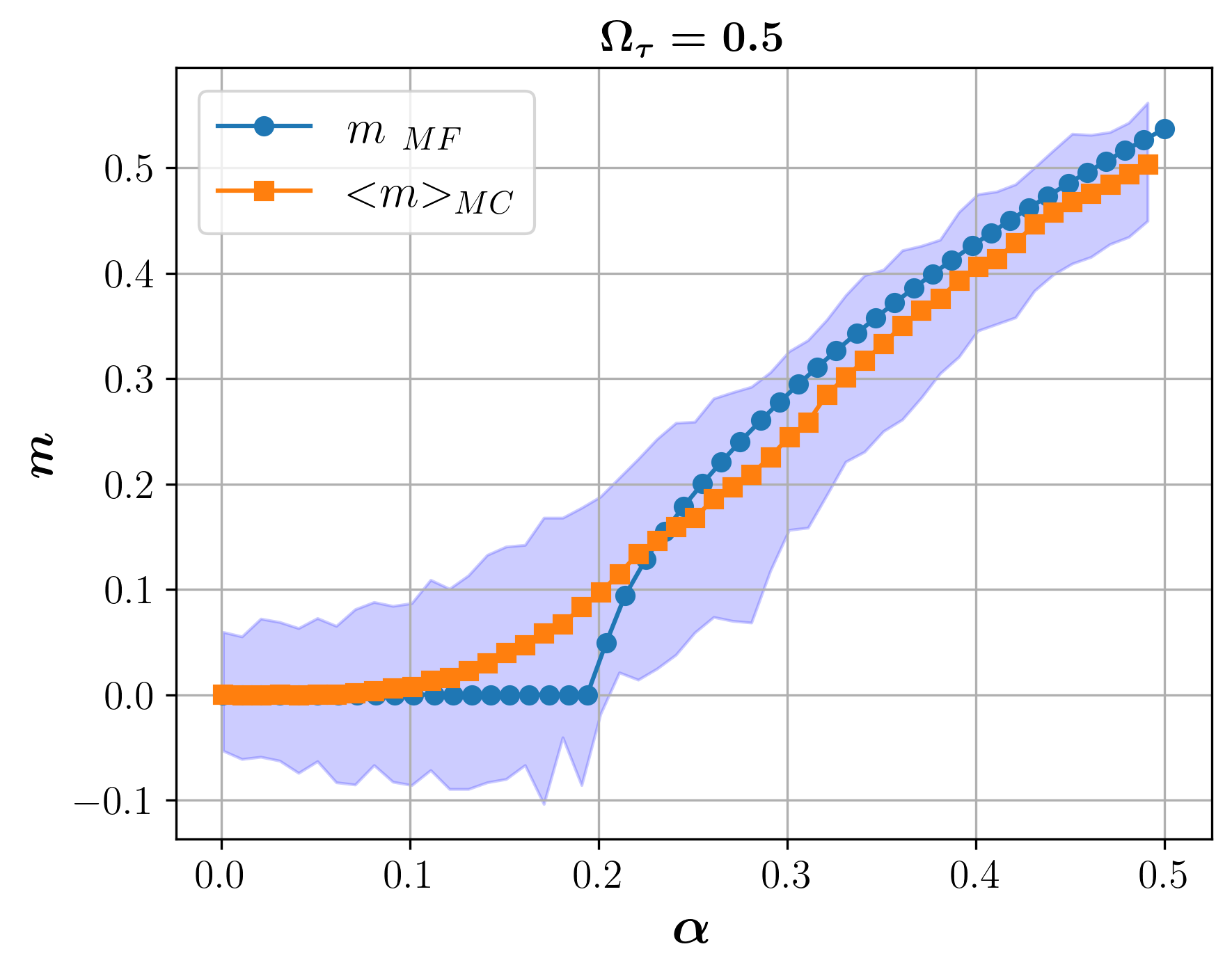}
\end{subfigure}%
\begin{subfigure}{.6\textwidth}
  \centering
\hspace{-16.0cm}\includegraphics[width=.75\linewidth]{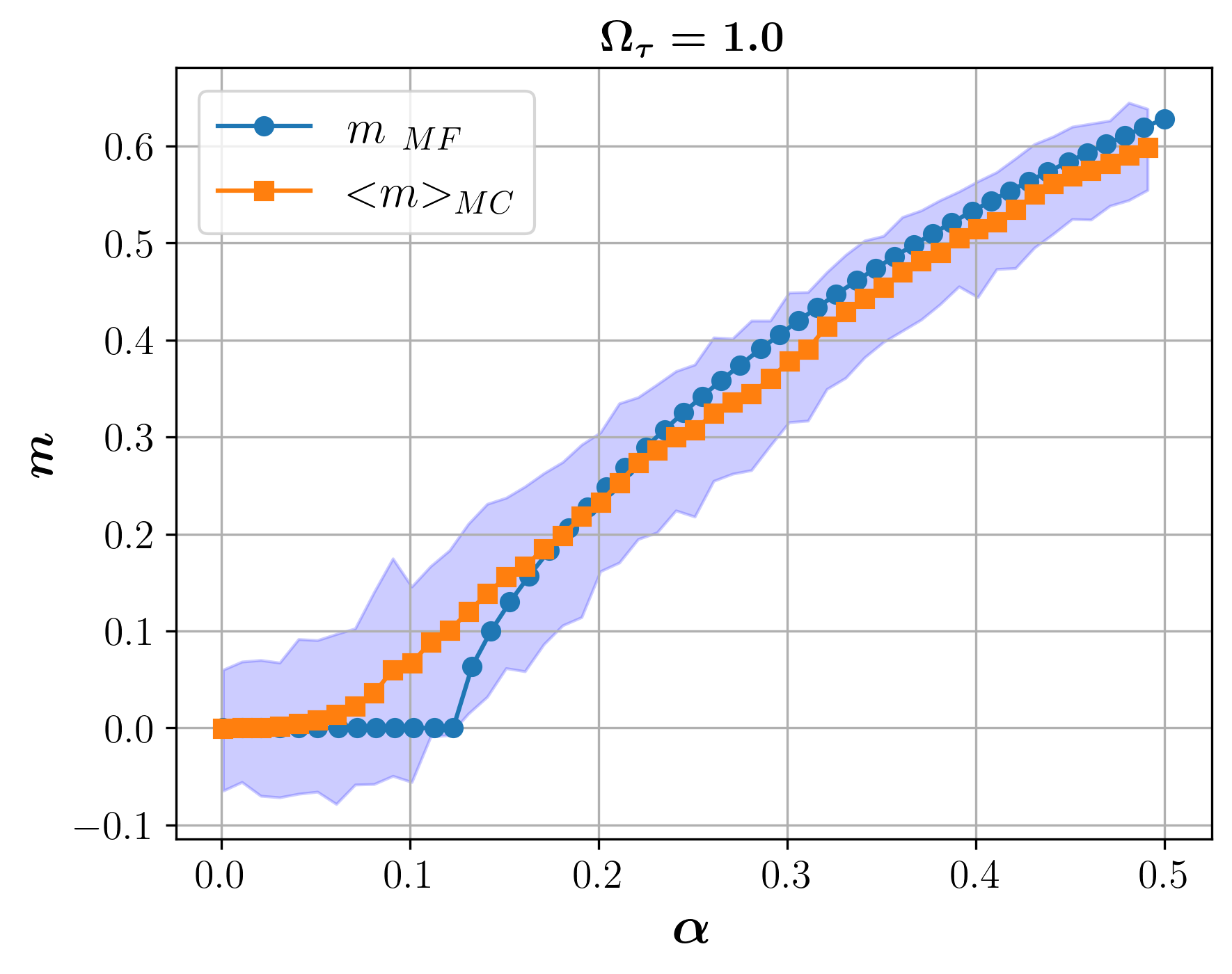}
\end{subfigure}
  \caption{ Agreement between theoretical replica analysis and Monte Carlo simulations. 
The magnetization computed from Eqs.~(\ref{eq:saddle_point_equations}--\ref{eq:self_overlaps}) is shown in blue, while the results from Monte Carlo simulations are shown in orange. 
The shaded area represents the range of values observed across runs. 
To compute the magnetization, we averaged over $N_s$ student pattern configurations $\xi$, sampled from the posterior distribution in Eq.~(\ref{eq:post_prob}). 
The figure compares the load $\alpha$ required to enter the signal retrieval (sR) phase at fixed student temperature $T = 1.5$, for different choices of hidden unit priors. }\label{fig: MCsimulations}
\end{figure}

Monte Carlo simulations shown in Fig.~(\ref{fig: MCsimulations}) are in agreement with the replica analysis and illustrate how the minimal load $\alpha$ required to enter the sR phase changes with the choice of prior.  Specifically, we fix the regularization strength for the student weights and set the temperature to $T = 1.5$, then evaluate the learning magnetization for different values of $\Omega_\tau = 0, 0.5, 1.0$.  As expected, Gaussian hidden units facilitate signal retrieval, as also evident in the phase diagram of Fig.~(\ref{fig: OTvariations}-\ref{fig: hopfieldTeacher OTvariations}).

\subsection{The role of the T-RBM priors}
By fixing the S-RBM architecture, one can instead investigate the impact of the data structure, specifically the T-RBM priors, on learning performance. For now, it is assumed that the examples consist of binary entries. From Eq. (\ref{eq: critical load}), an effect on both the minimum amount of data required to enter the sR phase and the retrieval temperature is expected.
\begin{figure}[H]
\centering
\begin{subfigure}{.6\textwidth}
  \centering
\hspace{-8.0cm}\includegraphics[width=.75\linewidth]{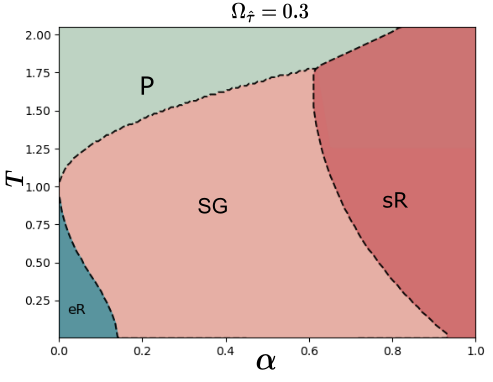}
\end{subfigure}%
\begin{subfigure}{.6\textwidth}
  \centering
\hspace{-12.0cm}\includegraphics[width=.75\linewidth]{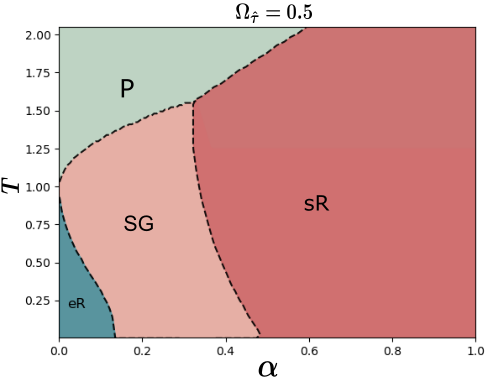}
\end{subfigure}%
\begin{subfigure}{.6\textwidth}
  \centering
\hspace{-16.0cm}\includegraphics[width=.75\linewidth]{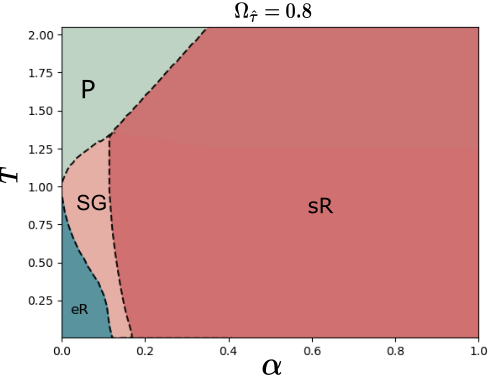}
\end{subfigure}
 \caption{Effect of the teacher architecture ($\hat\beta=0.8$) on the student's phase diagram. The signal's prior is defined by $\Omega_{\hat\xi}=0$. The analyzed student is working with a Hopfield architecture: $\Omega_{\xi} =0$,$\Omega_{\tau}=1$. The more $\hat{\tau}$ is gaussian the more easily the student retrieves the pattern $\bm{\hat\xi}$, in particular $P_{\text{triple}}$ decreases in both its coordinates.}
 \label{fig: parallel OTvariations}
\end{figure}
Fig. (\ref{fig: parallel OTvariations}) shows the effect of changing $\Omega_{\hat\tau}$. It is possible to appreciate the fact that the information present in the data increases notably the more the $\hat\tau$ prior becomes Gaussian. This can be seen by the reduced amount of generated samples we need to enter the sR regime (a similar effect of increasing $\hat\beta$ in \cite{alemanno2023hopfield}).

On the other hand, the critical temperature of the student at the triple point, $T_c$, is lowered. Nevertheless, if we fix an inference temperature, the student enters the sR regime more easily as $\Omega_{\hat\tau} \to 1$.
\begin{figure}[H]
\centering
\begin{subfigure}{.6\textwidth}
  \centering
\hspace{-8.0cm}\includegraphics[width=.75\linewidth]{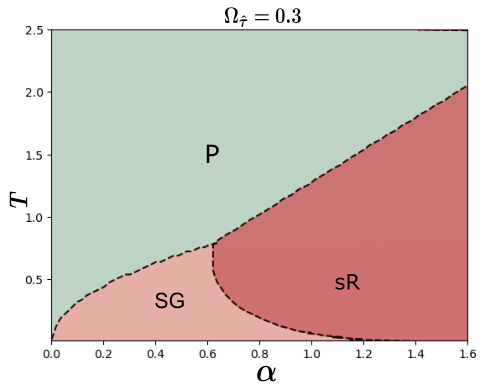}
\end{subfigure}%
\begin{subfigure}{.6\textwidth}
  \centering
\hspace{-12.0cm}\includegraphics[width=.75\linewidth]{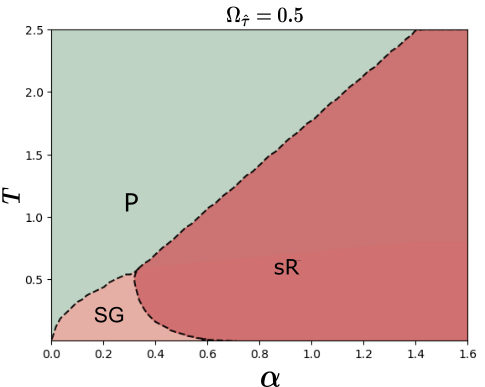}
\end{subfigure}%
\begin{subfigure}{.6\textwidth}
  \centering
\hspace{-16.0cm}\includegraphics[width=.75\linewidth]{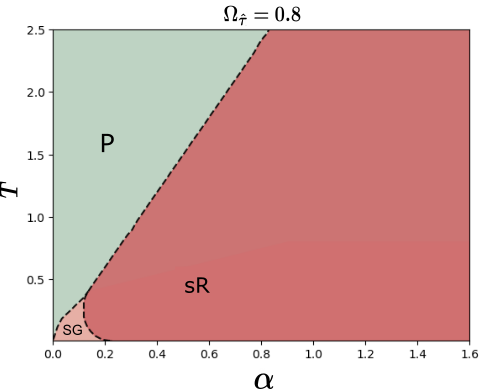}
\end{subfigure}
 \caption{Effect of the teacher ($\hat\beta=0.8$) architecture on a binary student's phase diagram. Since $\Omega_\tau=0$ we do not have example memorization. As in Fig.(\ref{fig: hopfieldTeacher OTvariations}) when $\Omega_{\hat{\tau}}\to1$ the S-RBM retrieval region becomes larger.
 }
 \label{fig: OT variation SBinary}
\end{figure}
Same effect is shown in Fig. (\ref{fig: OT variation SBinary}) where the student priors are binary and the teacher prior changes. We checked that the phase diagrams are not affected by $\bm{\hat\xi}$, as expected from the transition line expressions.
\begin{figure}[H]
\centering
\begin{subfigure}{.6\textwidth}
  \centering
\hspace{-8.0cm}\includegraphics[width=.75\linewidth]{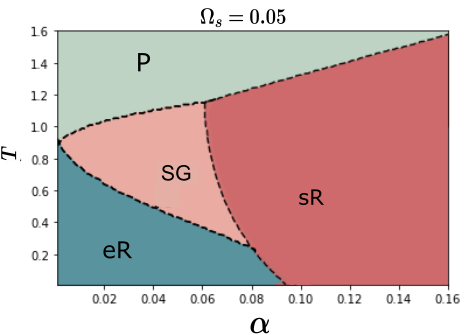}
\end{subfigure}%
\begin{subfigure}{.6\textwidth}
  \centering
\hspace{-12.0cm}\includegraphics[width=.75\linewidth]{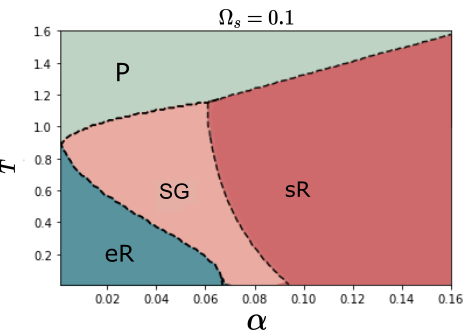}
\end{subfigure}%
\begin{subfigure}{.6\textwidth}
  \centering
\hspace{-16.0cm}\includegraphics[width=0.82\linewidth]{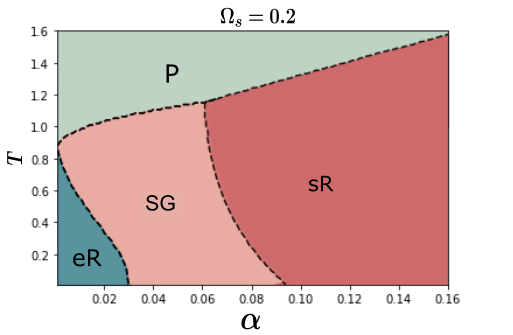}
\end{subfigure}
 \caption{Phase diagram of the teacher-student problem ($\hat\beta=0.8$) with $\bm \Omega =\{0,1,0,0.9, \Omega_s\}$, $\Omega_s\in\{0.005,0.1,0.2\}$. The effect of the manipulation of the data affects only the eR phase. The more the data have gaussian nature, the more the example memorization region is compressed.}
 \label{fig: Data prior variation}
\end{figure}
Finally in Fig. (\ref{fig: Data prior variation}) we explore the role of the dataset unit prior $\Omega_s$. The different values of $\Omega_s$ do not change the triple point $P_c$. As one can observe the only effect is the shift of the eR-SG transition line. This is related to the \textit{memorization} capability of the S-RBM and is in agreement with the phase diagrams  in \cite{barra2018phase}.  

\subsection{Spherical regularization}
\label{subsucsec: Spherical reg}

In the previous sections we have seen that the term $z(\bm\xi)^{-M}$ in the posterior plays the role of a regularization enforcing the Nishimori identity $d=1$. Without that regularization the model would be ill defined in the case of an architecture with both $\Omega_\xi$ and $\Omega_\tau$ different from zero. This is a typical problem with fully Gaussian disordered systems at low temperature \cite{barra2018phase,barra2012glassy,barra2014solvable,genovese2015legendre}.

\begin{figure}[H]
\centering
\begin{subfigure}{.6\textwidth}
  \centering
\hspace{-8.0cm}\includegraphics[width=.75\linewidth]{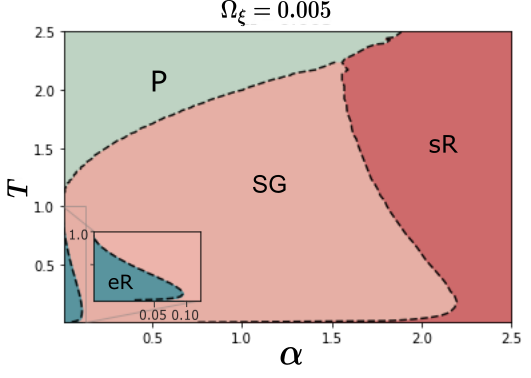}
\end{subfigure}%
\begin{subfigure}{.6\textwidth}
  \centering
\hspace{-12.0cm}\includegraphics[width=.75\linewidth]{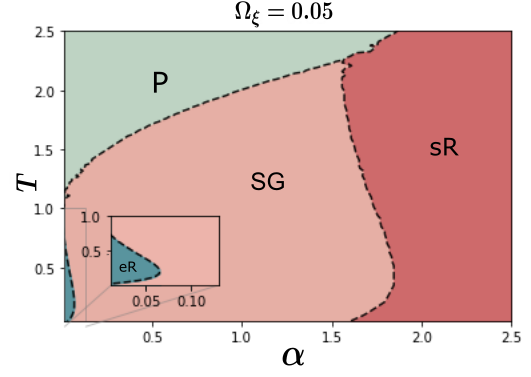}
\end{subfigure}%
\begin{subfigure}{.6\textwidth}
  \centering
\hspace{-16.0cm}\includegraphics[width=.75\linewidth]{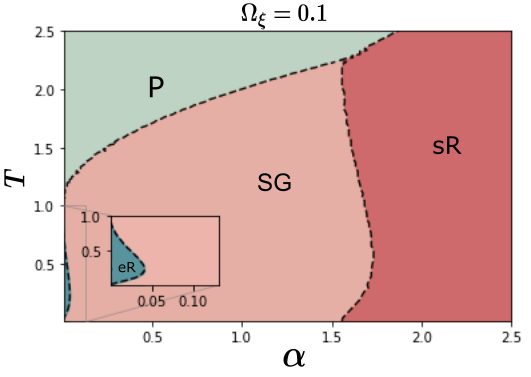}
\end{subfigure}
  \caption{ Spherical regularization  effect on the retrieval capability of the model (case with $\Omega_\tau =1$). Using binary data from a binary T-RBM ($\hat\beta=0.8$), it is possible to see that  $\alpha_c$ and $T_c$ do not move. The only effect from the changes of $\Omega_\xi$ is the compression of the $eR$ phase.  
  }
\label{fig: Soft prior variation}
\end{figure}

From a practical point of view is more convenient to replace that term directly with a spherical constraint 
$\delta \left( N - \sum_{i=1}^N \xi_i^2 \right)$. The computation leading to the set of saddle point Eqs. (\ref{eq:saddle_point_equations}-\ref{eq:self_overlaps}) can be exactly followed with the only addition of a lagrange multiplier $\omega$ in the gaussian weight for the $\xi$ variable, i.e. its effective distribution reads as
\begin{equation}
\label{eq: xi distribution with multiplier}
    \hspace{-1cm}P_{\Omega_{\xi}}(\xi) \EXP{ -\frac{\alpha \beta}{2} \left(\alpha\beta q_{\tau} - d_{\tau} + \omega \right)\xi^{2} + \xi\left(\alpha\sqrt{\hat{\beta}\beta} m_{\tau}\hat{\xi} + \sqrt{\alpha\beta q_{\tau}}z + \beta \Omega_{\tau}p s\right)}\,,
\end{equation}
where $\omega$ has to be fixed to obtain $d=1$. 
Making a comparison with Eq.(\ref{eq:mean_field_distribution1}) one can see that the correct choice is $\omega=\langle \tau^2\rangle_{\tau}$. From the computational point of view it is more convenient to adjust the Lagrange multiplier during the solution of the saddle point equations. In Fig.(\ref{fig: Soft prior variation}) it is shown the phase diagram where the priors of both the weights and the student hidden variables have Gaussian tails. The weights have been regularized with a spherical constraint. One can compare these results with the right panel of Fig.(\ref{fig: OTvariations}), where $\Omega_\xi =0$. It is possible to see the presence of the \textit{eR} phase, due to $\Omega_{\tau}=1$, that tends to quickly disappear as $\Omega_\xi\to 1$, in agreement with \cite{barra2018phase}. It is also observed a bending of the \textit{sR} region at low temperature that however is not particularly significant since that is the region where RSB corrections are expected. 
Finally, as expected,  the triple point position doesn't change.

\section{Conclusions}
In this paper, we investigate the learning performance of Restricted Boltzmann Machines (RBMs) trained on a noisy dataset in a teacher-student setting. We considered RBMs with different unit and weight priors, ranging from Gaussian to binary, and analyzed the generalization capability in both the Bayes optimal scenario—where the teacher and student share the same parameters—and in a more realistic mismatched case. We proved the existence of a well-defined critical dataset size below which learning is impossible, characterizing a triple point in the phase diagram where paramagnetic, spin glass, and signal retrieval phases intersect.

We show that continuous (Gaussian) hidden units help the machine to more easily enter the signal retrieval region, even though the critical size only depends on the properties of the dataset, i.e., the teacher RBM. At the same time, certain choices of priors favor the emergence of an example retrieval region where the machine learns by memorization without generalizing well. Notably, the posterior distribution approach for training includes automatic self-regularization of the weights, preventing parameter explosions and other typical convergence issues.
 By combining replica analysis with numerical simulations, we explicitly tested how the choice of priors for the hidden units affects learning performance, confirming that Gaussian priors lead to better signal retrieval with a smaller data requirement. Although the teacher’s parameters are unknown in practice, the mismatched analysis provides useful insight into which student architectures—such as those using Gaussian units—tend to be more robust across different dataset properties. In this sense, our framework can guide the identification of architectural choices that generalize well and help establish lower bounds on dataset size required for successful learning, independently of the precise nature of the data-generating process.\\
 A promising extension of this work would be to incorporate more structured data-generating processes, such as those involving correlated planted signals. Prior studies have explored such settings in specific regimes,  e.g. RBM with binary synapses in \cite{hou2019minimal,theriault2024modelling}, showing the occurrence of permutation symmetry breaking phase transitions. Similar transitions are observed in \cite{bachtis2024cascade}, where the training dynamics on a dataset generated from correlated patterns has been analyzed. Recent approaches \cite{xu2025learning} introduce a structure in the dataset by directly using a spiked covariance model as generating model for the examples. Extending our framework, which smoothly interpolates between Gaussian and binary priors, to these structured scenarios could reveal universal behaviors in feature-learning dynamics across the full class of RBMs. 
Such analysis would offer deeper insight into the interplay between data structure, noise, and architectural choices in shaping generalization performance in more realistic learning settings.

\appendix

\subfile{Generic_Prior_Equations.tex}

\subfile{Explicit_Generic_Eqs.tex}

\section*{Acknowledgement}
This work was partially supported by project SERICS (PE00000014) under the MUR National Recovery and Resilience Plan
funded by the European Union - NextGenerationEU. The work was also supported by the project PRIN22TANTARI "Statistical Mechanics of Learning Machines: from algorithmic and information-theoretical limits to new biologically inspired paradigms" 20229T9EAT – CUP J53D23003640001. DT also acknowledges GNFM-Indam.

\bibliographystyle{elsarticle-num}

\input{main.bbl}





\end{document}

%% file: Generic_Prior_Equations.tex
\section{Derivation of the RS equations}
\label{Sec:: Appendix A}
In this section we discuss the computation of the averaged partition function (\ref{eq: student partition function intro}) in the mismatch case with one single student pattern. 
The posterior of the student's (single) weight is 
\begin{eqnarray}
   P(\bm{\xi}|\bm{\mathcal{S}}) = Z^{-1}(\bm{\mathcal{S}})  z^{-M}(\bm\xi) P(\bm\xi) \mathbb{E}_{\bm\tau} \exp\left(\sqrt{\frac{\beta}{N}}\sum_{i=1}^N\sum_{a=1}^{M}  \xi_i s_i^a\tau^a \right)
\end{eqnarray}
then its partition function $Z(\bm{\mathcal{S}})$ can be written in the following form
\begin{eqnarray}
     Z(\bm{\mathcal{S}}) &=& \mathbb{E}_{\bm{\xi}} z^{-M}(\bm\xi)\mathbb{E}_{\bm{\tau}} \exp\left( \sqrt{\frac{\beta}{N}} \sum_{i=1}^{N} \sum_{a=1}^{M}\xi_i {s}_i^a \tau^a  \right)\,,
    \\
    z({\bm{\xi}})^M&=&\sum_{\bm{\mathcal{S}}}\mathbb{E}_{{\bm{\tau}}} \exp{\Big(\sqrt{\frac{{\beta}}{N}}\sum_{i=1}^{N}\sum_{a=1}^{M}{s}_{i}^{a}{\tau}^{a} {\xi}_{i} \Big)} \label{Aeq: zxi^M}\,.
\end{eqnarray}
Differently from the teacher, which generates the data at inverse temperature $\hat{\beta}<1$, the student's temperature range is $T\in [0,\infty]$. Due to this possibility, when $T<1$, $\bm\xi$ has the chance to be aligned with one of the data, say the first one $\bm{s}^{(1)}$. The alignment can be in principle valid also for the extensive hidden unit vector $\bm\tau$. It is measured with the two Mattis magnetizations
\begin{eqnarray}
\label{eq:magnetization_examples}
    p &=& \frac{1}{N} \sum_{i=1}^N \xi_i s_i^{(1)}\,, \\
    \label{eq:magnetization_examples_tau}
    p_{\tau} &=& \frac{1}{M} \sum_{a=1}^M \tau^a s_{(1)}^a \,, 
\end{eqnarray}
which allow to rewrite the previous partition function (discarding irrelevant non extensive terms) as
\begin{align*}
        Z(\bm{\mathcal{S}})
        = \mathbb{E}_{\bm\xi}  z^{-M}(\bm\xi)   
        \mathbb{E}_{\bm{\tau}}
        \exp \left( 
    M\sqrt{\frac{\beta }{N}} 
    p_{\tau}(\tau)\,\xi_{(1)}+
    N\sqrt{\frac{\beta }{N}} 
p(\xi)\,\tau^{(1)}
    +\sqrt{\frac{\beta}{N}}\sum_{i=2}^{N}\sum_{a=2}^{M}{s}_{i}^{a}\tau^{a}\xi_{i} 
    \right)\, .
\end{align*}

In spin-glass models with planted disorder we introduce $n$ independent replicas, then take the limit $N\to \infty$
\begin{align}
\label{Aeq:: [Z^n]}
\hspace{-1cm}
        Z^n(\bm{\mathcal{S}})
        = \mathbb{E}_{\{\bm{\xi}^\gamma\}} z(\{\bm{\xi}^\gamma\})^{-M}   
        \mathbb{E}_{\{\bm{\tau}^\gamma\}}
        \exp \Bigg( &
    M\sqrt{\frac{\beta }{N}} 
    \sum_{\gamma=1}^n p_{\tau}^{\gamma}(\tau)\xi_{(1)}^{\gamma}+
    N\sqrt{\frac{\beta }{N}} \sum_{\gamma=1}^n p^{\gamma}(\xi)\tau^{\gamma}_{(1)}\\
    &
    +\sqrt{\frac{\beta}{N}} \sum_{\gamma=1}^n \sum_{i=2}^{N}\sum_{a=2}^{M}{s}_{i}^{a}\tau^{a,\gamma}\xi_{i}^{\gamma} 
    \Bigg)\, .
\end{align}
The shorthand notation $\{\bm{\xi}^\gamma\}$ stands for the average over all the replicas indexed by $\gamma$: $\{\bm{\xi}^1,\ldots,\bm{\xi}^n\}$, same is done for $\{\bm{\tau}^\gamma\}$ .
Then the averaged partition function over the quenched disorder induced by the dataset is
\begin{align}
\left[Z^n\right]^{\bm{\mathcal{S}}}  = \sum_{\bm{\mathcal{S}}} P_{\hat{\beta}}(\bm{\mathcal{S}})\ Z^n(\bm{\mathcal{S}})
        &= \sum_{\bm{\mathcal{S}}}
        \,\mathbb{E}_{\hat{\bm{\xi}}}\mathbb{E}_{\hat{\bm{\tau}}}\ z(\hat{\bm{\xi}})^{-M}\  \EXP{\sqrt{\hat{\beta}/N} \sum_{ia}s_i^a \hat{\xi}_i \hat{\tau}_a} \ Z^n(\bm{\mathcal{S}}) \label{eq:: appendix Student Z}\,,\\
z(\hat{\bm{\xi}})^M&=\sum_{\bm{\mathcal{S}}}\mathbb{E}_{\hat{\bm{\tau}}} \exp{\Big(\sqrt{\frac{\hat{\beta}}{N}}\sum_{i=1}^{N}\sum_{a=1}^{M}{s}_{i}^{a}\hat{\tau}^{a} \hat{\xi}_{i} \Big)}\,.\label{eq:: appendix Teacher Z}
\end{align}
The presence of both $z(\hat{\bm{\xi}})$ and $z(\bm{\xi})$ make it diffucukt to compute $  \left[Z^n\right]^{\bm{\mathcal{S}}}$ as a straightforward extremization problem. A possible solution is the application of the saddle point method to both (\ref{eq:: appendix Teacher Z}) and (\ref{Aeq: zxi^M}). We can write them as an exponential function to be extremized over a set of parameters $\bm p_T$ and $\bm p_S$. Each set will depend respectively on $\bm{\hat{\xi}}$ and $\bm{\xi}$      

\begin{eqnarray}
\hspace{-1.2cm}
    z(\bm{\hat{\xi}})^M \approx  \EXP{-N\extr_{\bm p_T} \hat{\beta}f_T (\bm p_T(\bm{\hat{\xi}}))} \,, \  \    z(\bm{\xi})^M \approx \EXP{-N\extr_{\bm p_S} \beta  f_S(\bm p_S(\bm{\xi}) )} 
    \label{eq:: variational principle}
\end{eqnarray}
which, after the extremization, could be inserted in the exponential of (\ref{eq:: appendix Student Z}).

\subsubsection*{Teacher normalization}
In order to evaluate Eq.(\ref{eq:: appendix Teacher Z}) as a variational principle we start computing $\sum_{\bm{\mathcal{S}}}$ writing each $\bm{s}^a \in \bm{\mathcal{S}} $ in the interpolating form $\sqrt{\Omega_s} \bm{g}_s^a + \sqrt{\delta_s} \bm\epsilon_s^a$ for each $a = 1, \cdots, M$, obtaining
\begin{align}
    z(\hat{\bm\xi})^M= \int \prod_a \left[\mathcal{D}{\bm g}^{a} \right]\, \mathbb{E}_{\bm{\hat{\tau}}} \exp & \left( \sqrt{\frac{\hat\beta}{N}}\sum_{i=1}^{N} \sum_{a=1}^{M}\sqrt{\Omega_s}\,g_{s,i}^{a}\hat{\tau}^{a} \hat\xi_{i}\right)   \times \notag \\ 
    &\times \prod_{i,a}2 \cosh\left\{ \sqrt{\frac{\hat\beta}{N}}\sqrt{\delta_s}\hat{\tau}^{a} \hat\xi_{i}\right\} 
    \label{eq: A aprox teacher z 1} 
\end{align}
\begin{equation}
    \hspace{-2.2cm} \approx   \mathbb{E}_{\bm{\hat\tau}}\,\exp\left( \frac{\hat\beta}{2N}\sum_{i=1}^{N}\sum_{a=1}^{M}\left(\hat{\tau}_{a}\hat\xi_{i}\right)^{2}\right) \,,
    \label{eq: A aprox teacher z}
\end{equation}

where, in the first passage we already average over the set $\{ \bm\epsilon_s^a \}_{a=1,\cdots,M}$, while in the last one, we exploit $ \log \cosh(x) \approx \frac{x^2}{2} $ and $\Omega_s+\delta_s =1$. It is possible to use the Fourier representation of the Dirac delta function to enforce the relation 
\begin{eqnarray}
    \delta \left( d_{\hat{\tau}} - \frac{1}{M}\sum_a \hat\tau_a^2 \right)&=& \frac{1}{2\pi}\int d\hat d_{\hat\tau}\, \EXP{-M\hat d_{\hat{\tau}}\left(d_{\hat{\tau}}-\frac{1}{M}\sum_a \hat{\tau}_{a}^2  \right)}\,,
\end{eqnarray}
where we refer at $ d_{\hat\tau}$ as the self-overlap for the teacher hidden variables. Using the following expression for the density of states of the $\hat\tau$ configurations 
\begin{eqnarray}
    \hspace{-1cm}\mathcal{D}(\hat d_{\hat\tau}
    ,d_{\hat\tau}) & = \sum_{\bm{\hat\tau}}  \delta \left( d_{\hat{\tau}} - \frac{1}{M}\sum_a \hat\tau_a^2 \right) \propto \sum_{\bm{\hat\tau}} \int d\hat d_{\hat\tau}
     \; \EXP{-M\hat d_{\hat{\tau}}\left(d_{\hat{\tau}}-\frac{1}{M}\sum_a \hat{\tau}_{a}^2  \right)} \,,
\end{eqnarray}
we can rewrite (\ref{eq: A aprox teacher z}) 
\begin{eqnarray}
    z(\hat{\bm\xi})^M &=& \int  dd_{\hat\tau} \; \mathcal{D}(\hat d_{\hat\tau}
    ,d_{\hat\tau}) \exp \left( \frac{\alpha \hat\beta}{2} d_{\hat\tau}\sum_i \hat\xi_i^2 \right) \\
    &\approx&  \EXP{-N\hat\beta\operatorname{Extr}_{\bm p_T}  f(\bm p_T)}\,,
    \label{eq:: App variational principle1}
\end{eqnarray}
which is nothing but the variational principle in (\ref{eq:: variational principle}) with $\bm p_T = \{\hat d_{\hat\tau},d_{\hat\tau}\}$. Using 
$d_{\hat\xi}( \hat{\bm\xi})=\frac{1}{N}\sum_i \hat\xi_i^2$ as well, for the self overlap of the teacher pattern, the  function to be extremized in (\ref{eq:: App variational principle1}) is 
 
\begin{eqnarray}
    -\hat\beta f(\bm p_T) = \alpha\left(-d_{\hat{\tau}}\hat{d}_{\hat{\tau}}+\frac{\hat\beta}{2}d_{\hat\xi}(\hat{\bm\xi})\hat{d}_{\hat{\tau}}+\log\mathbb{E}_{\hat{\tau}}\EXP{\hat{d}_{\hat{\tau}}\hat{\tau}^{2}}\right)
\end{eqnarray}
reaching its minimum at
\begin{eqnarray}
\label{eq: A teacher den saddle point}
    \hat{d}_{\hat{\tau}}=\frac{\hat\beta}{2}d_{\hat \xi}(\hat{\bm\xi})\,, \qquad \qquad d_{\hat{\tau}}=\langle \hat{\tau}^2 \rangle_{\hat\tau} = \frac{\mathbb{E}_{\hat\tau} \hat\tau^2 \EXP{\frac{\hat\beta}{2}    d_{\hat \xi}(\hat{\bm\xi}) \hat\tau^{2} } }{\mathbb{E}_{\hat\tau} \EXP{\frac{\hat\beta}{2} d_{\hat \xi}(\hat{\bm\xi}) \hat\tau^{2}}}\,.
\end{eqnarray}
In view of the above, the averaged partition function (\ref{eq:: appendix Student Z}) can be approximated as
\begin{eqnarray}
    \hspace{-1.5cm}\left[Z^n\right]^{\bm{\mathcal{S}}}  
        &= &\sum_{\bm{\mathcal{S}}}
        \,\mathbb{E}_{\hat{\bm{\xi}}}\mathbb{E}_{\hat{\bm{\tau}}} \EXP{N\hat\beta \extr_{\bm p_T} f(\bm p_T)} \  \EXP{\sqrt{\hat{\beta}/N} \sum_{ia}s_i^a \hat{\xi}_i \hat{\tau}_a}  Z^n(\bm{\mathcal{S}}) 
\end{eqnarray}

\subsubsection*{Student normalization}
The same problem of the teacher normalization term  affects Eq.(\ref{Aeq:: [Z^n]}), due to the presence of $z(\{\bm\xi^{\gamma}\})$. As already stated in Eqs.(\ref{eq:: variational principle}) we want to write it in a variational principle manner. The steps to do that are the same of the one used for (\ref{eq: A aprox teacher z 1}-\ref{eq: A teacher den saddle point}).
We start computing the sum over the examples using their interpolated form, then we impose the self overlap with the Dirac delta
\begin{align}
    \hspace{-0.7cm}z(\{\bm\xi^{\gamma}\})^M = \int \prod_{a,\gamma}\left[\mathcal{D}\boldsymbol{g}^{a \gamma}\right]\,\mathbb{E}_{\{\bm\tau^\gamma\}}\exp& \left( \sqrt{\frac{\beta}{N}}\sum_{\gamma,i,a}\sqrt{\Omega_s}\,g_{s,i}^{a\gamma}{\tau}_{a}^{\gamma} \xi_{i}^\gamma \right)  \times
    \notag\\
    &\times \prod_{i,a,\gamma} 2 \cosh\left\{ \sqrt{\frac{\beta}{N}}\sqrt{\delta_s}\tau_{a}^\gamma \xi_{i}^\gamma \right\} 
    \label{eq: A aprox student z 1}
    \end{align}
    
    \begin{align}
  &\approx   \mathbb{E}_{\{\bm\tau^\gamma\}}\,\exp\left( \frac{\beta}{2N}\sum_{\gamma=1}^{n}\sum_{i=1}^{N}\sum_{a=1}^{M}\left(\tau_{a}^\gamma \xi_{i}^\gamma \right)^{2}\right) \,,
    \label{eq: A aprox student z} \\
    &\approx \int \prod_\gamma  
    \, dd_{\tau}^\gamma  \,\mathcal{D}(\bm{\hat d_{\tau}}
    ,\bm{d_{\tau}}) \exp \left( \frac{\alpha \beta}{2}\sum_\gamma d_{\tau}^\gamma d^\gamma ( {\bm\xi})\right) \\
    &\approx  \EXP{-N\beta\operatorname{Extr}_{\bm p_S}  f(\bm p_S)}\,,
    \end{align}

with density of states and self overlap for every student replica as 
\begin{eqnarray}
    \mathcal{D}(\bm{\hat d_{\tau}}
    ,\bm{d_{\tau}}) & = & \sum_{\{\bm\tau^\gamma \}}  \delta \left( \sum_\gamma d_{{\tau}}^\gamma  - \frac{1}{M}\sum_{a,\gamma} (\tau_a^\gamma)^2 \right) \\
    &\propto&
\sum_{\{\bm\tau^\gamma \}} \int \prod_\gamma  d\hat d_{\tau}^\gamma \, \EXP{-M \sum_\gamma \hat d_{\tau}^\gamma\left(d_{\tau}^\gamma -\frac{1}{M}\sum_a (\tau_{a}^\gamma)^2  \right)} \,,
    \notag
    \end{eqnarray}
\begin{equation*}
    d^\gamma( {\bm\xi}) = \frac{1}{N}\sum_i (\xi_i^\gamma)^2 \,.
\end{equation*}
Finally we can rewrite Eq.(\ref{Aeq:: [Z^n]}) as
\begin{eqnarray}
Z^n(\bm{\mathcal{S}})
        = \mathbb{E}_{\{\bm{\xi}^\gamma\}}   
        \mathbb{E}_{\{\bm{\tau}^\gamma\}}
        \exp &\Bigg( 
    M\sqrt{\frac{\beta }{N}} &
    \sum_{\gamma=1}^n p_{\tau}^{\gamma}(\tau)\xi_{(1)}^{\gamma}+
    N\sqrt{\frac{\beta }{N}} \sum_{\gamma=1}^n p^{\gamma}(\xi)\tau^{\gamma}_{(1)}+\\
    &
    +&\sqrt{\frac{\beta}{N}} \sum_{\gamma=1}^n \sum_{i=2}^{N}\sum_{a=2}^{M}{s}_{i}^{a}\tau^{a,\gamma}\xi_{i}^{\gamma} 
    +N\beta \extr_{\bm p_S} f_S(\bm p_S)  \Bigg)
    \notag \, .
\end{eqnarray}
The function $f_S(\bm p_S)$ together with the values of $\bm p_S$ that minimize it are
\begin{equation}
    \hspace{-0.7cm}-\beta f_S (\bm p_S) =  -\alpha\sum_\gamma d_{{\tau}}^\gamma \hat{d}_{{\tau}}^\gamma +\frac{\alpha \beta}{2}\sum_\gamma d^\gamma ({\bm\xi}){d}_{{\tau}}^\gamma +\frac{1}{N} \log\mathbb{E}_{\{\tau^\gamma_a \}}\EXP{\sum_{\gamma,a} \hat{d}_{{\tau}}^\gamma \sum_a (\tau^\gamma_a )^{2}}
\end{equation}
\begin{equation}
\label{A student denominator}
  \hat{d}_{{\tau}}^\gamma=\frac{\beta}{2}d^\gamma({\bm\xi}) \,, \qquad d^\gamma _{\hat{\tau}}=\langle {(\tau^\gamma)}^2 \rangle_\tau = \frac{\mathbb{E}_{\tau^\gamma} (\tau^\gamma)^2 \,\EXP{\frac{\beta}{2}  d^\gamma(\bm\xi)(\tau^{\gamma})^2} }{\mathbb{E}_{\tau^\gamma}  \EXP{\frac{\beta}{2}  d^\gamma(\bm\xi) (\tau^{\gamma})^2}}\,.
\end{equation}

\subsubsection*{RS free energy and saddle point equations}
Based on the preceding sections, we can rewrite the averaged partition function as 
\begin{align}
   \label{eq:: appendix Student Z 11}\left[Z^n\right]^{\bm{\mathcal{S}}}  =&\sum_{\{\xi_{(1)}^{\gamma}\}} \mathbb{E}_{\{\tau_{(1)}^{\gamma}\}}
        \sum_{\bm{s},\,\tilde{\bm{s}}}\exp{\left( M\sqrt{\frac{\beta }{N}} 
    p_{\tau}^{\gamma}(\tau)\xi_{(1)}^{\gamma}+
    N\sqrt{\frac{\beta }{N}} 
p(\xi)^{\gamma}\tau^{\gamma}_{(1)}
 \right)} \times \\
\label{eq:: appendix Student Z 12} 
        \times &\mathbb{E}_{\hat{\bm{\xi}}}\mathbb{E}_{\hat{\bm{\tau}}}
\sum_{\bm{\mathcal{S}}}\exp \left( N\hat\beta \extr_{\bm p_T} f_T(\bm p_T)  +\sqrt{\frac{\hat{\beta}}{N}}\sum_{i=2}^{N}\sum_{a=2}^{M}{s}_{i}^{a}\hat{\tau}^{a}\hat{\xi}_{i} \right) \times \\
\label{eq:: appendix Student Z 13}
\times& \sum_{\{\bm{\xi}^{\gamma}\}}  \mathbb{E}_{\{\bm{\tau}^\gamma\}}\; \exp\left( N\beta \extr_{\bm p_S}f_S(\bm p_S) +\sqrt{\frac{\beta}{N}}\sum_{i=2}^{N}\sum_{a=2}^M \sum_{\gamma=1}^{n}{s}_{i}^{a}\tau^{a,\gamma}\xi_{i}^{\gamma} 
        \right) \;.
\end{align}
By incorporating the variational principle approximation (\ref{eq:: variational principle}) into (\ref{Aeq:: [Z^n]}) and (\ref{eq:: appendix Student Z}), we account for both $z(\hat{\bm\xi})^{-M}$ and $z({\{\bm\xi^\gamma}\})^{-M}$. Additionally we isolate the average over the first example as described in  (\ref{eq:magnetization_examples}, \ref{eq:magnetization_examples_tau}), denoting $\bm s = (s_1^{(1)},\cdots,s_N^{(1)})$ and $\tilde {\bm s} = (s^1_{(1)},\cdots,s^M_{(1)})$.  
 Starting with the first term (\ref{eq:: appendix Student Z 11}) of the quenched free energy, this term can be transformed by constraining the overlap with the examples $p_{\tau}^{\gamma}$ and $p^{\gamma}$
\begin{align}
& \int \, \prod_\gamma dp^\gamma\,\prod_\gamma dp_{\tau}^\gamma \,\mathcal{D}(\bm{p},\bm{p}_{\tau})
    \sum_{\{\xi_{(1)}^{\gamma}\}} \mathbb{E}_{\{\tau_{(1)}^{\gamma}\}} \exp{\left( M\sqrt{\frac{\beta }{N}} \sum_{\gamma}
    p_{\tau}^{\gamma}\xi_{(1)}^{\gamma}+
    N\sqrt{\frac{\beta }{N}}\sum_{\gamma} 
    p^{\gamma}\tau^{\gamma}_{(1)}
 \right)} \notag
 \\ = & \int \, \prod_\gamma dp^\gamma\,\prod_\gamma dp_{\tau}^\gamma \,\mathcal{D}(\bm{p},\bm{p}_{\tau})
        \exp{\left( \frac{N \beta }{2}  \alpha^2 \Omega \sum_{\gamma}
    (p_{\tau}^{\gamma})^2 +
    \frac{N \beta }{2} \Omega_{\tau} \sum_{\gamma} 
    (p^{\gamma})^2
 \right)} \label{eq:: appendix phase space counting p}\,,
\end{align}
where we introduced the notation for the density of states for the configurations $\bm{s}$ and $\bm{\Tilde{s}}$
\begin{align*}
    \mathcal{D}(\bm{p},\bm{p}_{\tau}) & = \sum_{\bm{s},\,\Tilde{\bm{s}}} \prod_{\gamma} \delta \left(  p^{\gamma}-\frac{1}{N}\sum_{i} s_i \xi_i \right) \prod_{\gamma} \delta \left(  p^{\gamma}_{\tau}-\frac{1}{N}\sum_{a} \Tilde{s}^{a}\tau^a \right) \\
     & \propto \sum_{\bm{s},\,\Tilde{\bm{s}}} \int \, \prod_{\gamma} d\hat{p}^{\gamma} \prod_{\gamma} d\hat{p}_{\tau}^{\gamma} \exp \left( -N\sum_{\gamma}\hat{p}_{\gamma}\left(p_{\gamma}- \frac{1}{N}\sum_i s_i \xi^{\gamma}_i \right) \right) \times
     \\
     &\hspace{3.7cm} \times \exp \left(-M\sum_{\gamma}\hat{p}^{\gamma}_{\tau}\left(p^{\gamma}_{\tau}- \frac{1}{M}\sum_{a} \Tilde{s}_a \tau^{\gamma}_a \right)  \right) \;.
\end{align*}
Next, we turn our attention to the second and third terms, (\ref{eq:: appendix Student Z 12}) and (\ref{eq:: appendix Student Z 13}), which emphasize the data-dependent components. Since these terms are already linear in $s_i^a$ we can already compute the average over the examples
\begin{align}
& \sum_{\bm{\mathcal{S}}}   \sum_{\bm{\xi}^1,\ldots,\bm{\xi}^n}  \mathbb{E}_{\{\bm{\tau}^\gamma\}}\mathbb{E}_{\hat{\bm{\xi}}}\;\mathbb{E}_{\hat{\bm{\tau}}} \EXP{ \sum_{i,a} {s}_{i}^{a} \left(\sqrt{\frac{\hat{\beta}}{N}}\hat{\tau}_{a}\hat{\xi}_{i} +\sqrt{\frac{\beta}{N}} \sum_{\gamma}\tau_{a}^{\gamma}\xi_{i}^{\gamma}\right) } =\\
&=\sum_{\bm{\xi}^1,\ldots,\bm{\xi}^n}  \mathbb{E}_{\{\bm{\tau}^\gamma\}}\mathbb{E}_{\hat{\bm{\xi}}}\;\mathbb{E}_{\hat{\bm{\tau}}}\EXP{\sum_{i,a} \ln \cosh(\left(\sqrt{\frac{\hat{\beta}}{N}}\hat{\tau}_{a}\hat{\xi}_{i} +\sqrt{\frac{\beta}{N}} \sum_{\gamma}\tau_{a}^{\gamma}\xi_{i}^{\gamma}\right)} \notag \\
   &\approx  \sum_{\bm{\xi}^1,\ldots,\bm{\xi}^n}  \mathbb{E}_{\{\bm{\tau}^\gamma\}}\;\mathbb{E}_{\hat{\bm{\xi}}}\;\mathbb{E}_{\hat{\bm{\tau}}} \,\EXP{ \frac{M\hat{\beta}}{2} d_{\hat\tau} d_{\hat\xi} +M \beta \sum_{\gamma < \gamma'} q_{\tau}^{\gamma \gamma'}q^{\gamma \gamma'} +\frac{M\beta}{2}\sum_{\gamma} d^{\gamma}_{\tau}d^{\gamma} +M\sqrt{\hat{\beta}\beta} \sum_{\gamma} m_{\tau}^{\gamma}m^{\gamma} } \;,
\end{align}
in the last line we expanded the $\log \cosh$ and introduced the order parameters
\begin{align*}
     m^{\gamma} &= \sum_{i}\frac{\hat{\xi}_i\xi_i}{N}, \qquad m^{\gamma}_{\tau}= \sum_{a} \frac{\hat{\tau}_a\tau_a}{M} \,,\\
    d^{\gamma}&= \sum_i \frac{(\xi_i^{\gamma})^2}{N} \qquad
    d^{\gamma}_{\tau}= \sum_a \frac{(\tau_a^{\gamma})^2}{M}\,,
     \\
     q^{\gamma \gamma'} & = \sum_{i}\frac{\xi_i^{\gamma}\,\xi_i^{\gamma'}}{N} \qquad
    q^{\gamma \gamma'}_{\tau}= \sum_{a}\frac{\tau_a^{\gamma}\,\tau_a^{\gamma'}}{M}\,, \\
    d_{\hat\xi} &= \sum_i \frac{\hat{\xi}_i^2}{N} \qquad
    d_{\hat\tau}= \sum_a \frac{\hat{\tau}_a^2}{M}\,.
\end{align*}
Consequently, we can fix the values of these parameters, as previously done in (\ref{eq:: appendix phase space counting p}), and incorporate the corresponding densities of states $\mathcal{D}(\bm \Lambda, \bm \Lambda_\tau):=\mathcal{D}(\bm{m},\bm{q},\bm{d})\,\mathcal{D}(\bm{m}_{\tau},\bm{q}_{\tau},\bm{d}_{\tau})\,$\\
$\mathcal{D}(d_{\hat\xi},d_{\hat\tau})$. 
Both the densities affect the extremized functions $f_T$ and $f_S$, as their parameters depend respectively on $d_{\hat\xi}$ and $d^\gamma$. For brevity of notation we defined the sets of order parameters: $\bm\Lambda:=\{ \bm{p},\bm{m},\bm{q},\bm{d},d_{\hat\xi
}\}$ and $\bm\Lambda_{\tau}:= \{\bm{p}_{\tau}, \bm{m}_{\tau},\bm{q}_{\tau},\bm{d}_{\tau},d_{\hat{\tau}}\}$ and their conjugated ones $\hat{\bm\Lambda}$, $\bm{\hat{\Lambda}}_\tau$, obtaining the following form of the averaged partition function
\small{
\begin{align}
    \label{A_eq::  final partition function}\hspace*{-1.2cm}\left[Z^n\right]^{\bm{\mathcal{S}}}  = \int d\,\bm\Lambda d\,\bm\Lambda_{\tau} \mathcal{D}(\bm\Lambda,\bm{\Lambda}_{\tau}) \exp & \Big( 
    \frac{M\hat{\beta}}{2}d_{\hat\tau} d_\xi +M \beta \sum_{\gamma < \gamma'} q_{\tau}^{\gamma \gamma'}q^{\gamma \gamma'}+ \frac{N \beta }{2}  \alpha^2 \Omega \sum_{\gamma}
    (p_{\tau}^{\gamma})^2 +
    \frac{N \beta }{2} \Omega_{\tau} \sum_{\gamma} 
    (p^{\gamma})^2 +\\
    +&\frac{M\beta}{2}\sum_{\gamma} d^{\gamma}_{\tau}d^{\gamma} +M \sqrt{\hat{\beta}\beta} \sum_{\gamma} m_{\tau}^{\gamma}m^{\gamma} +N \hat{\beta}\extr_{\bm p_T} f_T(\bm p_T) +N \beta\extr_{\bm p_S} f_S(\bm p_S) 
 \Big)  \notag\,.
\end{align}
}
At this point we use the Laplace method on the averaged partition function (\ref{A_eq::  final partition function}). Remembering the replica trick we have 
\begin{align*}
      -\beta f(\beta,\hat{\beta},\alpha) =& \lim_{N \to \infty} \frac{1}{N} \left[ \ln Z \right]^{\bm{\mathcal{S}}} = \lim_{\substack{N \to \infty \\ n\to 0}}\frac{ \ln [Z^n]^{\bm{\mathcal{S}}}}{Nn} 
\end{align*}
\begin{align}
\label{eq:: Extr free energy}
          -\beta f 
 \approx   &\lim_{\substack{N \to \infty \\ n\to 0}} \frac{1}{N n} \ln \Bigg( \EXP{N\extr_{\substack{\scriptstyle \bm\Lambda, \bm{\Lambda}_{\tau}\\\scriptstyle \hat{\bm\Lambda}, \hat{\bm{\Lambda}}_{\tau}}} \hat f(\bm\Lambda, \bm{\Lambda}_{\tau},\hat{\bm\Lambda}, \hat{\bm{\Lambda}}_{\tau})} \Bigg) \approx  \lim_{n\to 0} \frac{1}{ n} \extr_{\substack{\scriptstyle \bm\Lambda, \bm{\Lambda}_{\tau}\\\scriptstyle \hat{\bm\Lambda}, \hat{\bm{\Lambda}}_{\tau}}} \hat f(\bm\Lambda, \bm{\Lambda}_{\tau},\hat{\bm\Lambda}, \hat{\bm{\Lambda}}_{\tau})
\end{align}\\
\begin{align}
\label{Aeq: replica free energy}
\hspace{-0.5cm}-\beta \hat f(\bm\Lambda, \bm{\Lambda}_{\tau},\hat{\bm\Lambda}, \hat{\bm{\Lambda}}_{\tau}) & = 
    \frac{\alpha \hat{\beta}}{2}d_{\hat\tau} d_{\hat\xi} +\alpha \beta \sum_{\gamma < \gamma'} q_{\tau}^{\gamma \gamma'}q^{\gamma \gamma'}+ \frac{ \beta }{2}  \alpha^2 \Omega_\xi \sum_{\gamma}
    (p_{\tau}^{\gamma})^2 +
    \frac{\beta }{2} \Omega_{\tau} \sum_{\gamma} 
    (p^{\gamma})^2 +  \\
    &+\frac{\alpha \beta}{2}\sum_{\gamma} d^{\gamma}_{\tau}d^{\gamma} +\alpha \sqrt{\hat{\beta}\beta} \sum_{\gamma} m_{\tau}^{\gamma}m^{\gamma} + \hat{\beta}\extr_{\bm p_T} f_T(\bm p_T) +  \notag \\
    & -\hat{d}_{\hat\xi} d_{\hat\xi} -\sum_\gamma \hat{p}^\gamma p^\gamma -\sum_{\gamma<\gamma'}\hat{q}^{\gamma \gamma'} q^{\gamma \gamma'}-\sum_\gamma \hat{m}^\gamma m^\gamma + \beta\extr_{\bm p_S} f_S(\bm p_S)+
    \notag \\
    & -\alpha \hat{d}_{\hat\tau} d_{\hat\tau} - \alpha \sum_\gamma \hat{p}_\tau ^\gamma p_\tau ^\gamma - \alpha \sum_{\gamma<\gamma'}\hat{q}_\tau ^{\gamma \gamma'} q_\tau ^{\gamma \gamma'}- \alpha \sum_\gamma \hat{m}_\tau ^\gamma m_\tau ^\gamma +
    \notag \\
    &\hspace{-3cm}+ \alpha \log \mathbb{E}_{\hat{\tau}} \mathbb{E}_{\tilde s} \mathbb{E}_{\{\tau^{\gamma}\}} \exp{\left(  \hat{d}_{\hat\tau}  \,\hat{\tau}^2 +  \frac{1}{2}   \sum_{\gamma \gamma'} \hat{q}_{\tau}^{\gamma \gamma'} \tau^{\gamma} \tau^{\gamma'}  -  \frac{1}{2} \sum_{ \gamma} \hat{q}_{\tau}^{\gamma \gamma}(\tau^{\gamma})^2 +  \sum_{ \gamma} \hat{m}_{\tau}^\gamma \tau^{\gamma} \hat{\tau} + \sum_{ \gamma} \hat{d}_{\tau}^\gamma (\tau^{\gamma})^2 +\sum_\gamma \tilde s p_\tau^\gamma \tau^\gamma \right)}+
    \notag \\
    &\hspace{-3cm}+ \log \mathbb{E}_{\hat{\xi}}\mathbb{E}_{s} \mathbb{E}_{\{{\xi}^{\gamma}\}} \exp{\left(  \hat{d}_{\hat\xi} \,\hat{\xi}^2 +  \frac{1}{2}  \sum_{\gamma \gamma'} \hat{q}^{\gamma \gamma'} \xi^{\gamma} \xi^{\gamma'}  -  \frac{1}{2} \sum_{ \gamma} \hat{q}^{\gamma \gamma}(\xi^{\gamma})^2 +  \sum_{ \gamma} \hat{m}^\gamma \xi^{\gamma} \hat{\xi} +  \sum_{\gamma} \hat{d}^\gamma (\xi^{\gamma})^2 +\sum_{ \gamma} s \,\hat{p}^\gamma \xi^\gamma \right)} \,,\notag
\end{align}
where, in the last two lines, we just collect the linearities in $a,i$ inside the exponentials of $\mathcal{D}(\bm \Lambda, \bm \Lambda_\tau)$ and substitute them with, respectively, $N$ and $M$ times the averages on one representative. As an example
\begin{align*}
    \log \mathbb{E}_{\hat{\bm\xi}}\mathbb{E}_{\bm s} \mathbb{E}_{\{{\bm\xi}^{\gamma}\}} &\exp\Bigg(  \hat{d} _{\hat\xi}  \sum_i \hat{\xi}^2_i +  \frac{1}{2}  \sum_{\gamma \gamma'} \hat{q}^{\gamma \gamma'} \sum_i \xi^{\gamma}_i \xi^{\gamma'}_i \! \!-\!  \frac{1}{2} \sum_{ \gamma} \hat{q}^{\gamma \gamma} \sum_i(\xi^{\gamma}_i)^2 + \\& +\sum_{ \gamma} \hat{m}^\gamma \sum_i \xi^{\gamma}_i \hat{\xi}_i +  \sum_{\gamma} \hat{d}^\gamma \sum_i(\xi^{\gamma}_i)^2 +\sum_{ i \gamma} s_i \,\hat{p}^\gamma \xi^\gamma_i \Bigg) =\\ 
\end{align*}
\vspace{-1.5cm}
\begin{equation*}
    \hspace{-1cm} =\log \left[ \mathbb{E}_{\hat{\xi}}\mathbb{E}_{s} \mathbb{E}_{\{{\xi}^{\gamma}\}} \exp{\left(  \hat{d}_{\hat\xi} \hat{\xi}^2 +  \frac{1}{2}  \sum_{\gamma \gamma'} \hat{q}^{\gamma \gamma'} \xi^{\gamma} \xi^{\gamma'}  -  \frac{1}{2} \sum_{ \gamma} \hat{q}^{\gamma \gamma}(\xi^{\gamma})^2 +  \sum_{ \gamma} \hat{m}^\gamma \xi^{\gamma} \hat{\xi} +  \sum_{\gamma} \hat{d}^\gamma (\xi^{\gamma})^2 +\sum_{ \gamma} s \,\hat{p}^\gamma \xi^\gamma \right)}\right]^N \,,
\end{equation*}

then, using the properties of the $\log$ function, one can express it as the extensive term of the last line of (\ref{Aeq: replica free energy}).\\
From now on we assume the Replica symmetric (RS) ansatz for all the order parameters: $q^{\gamma \gamma'}=q, m^\gamma=m, p^\gamma = p, d^\gamma =d,\ \forall \gamma,\gamma'=1,\ldots,n$, same is valid also for their $\tau$ version and their conjugated. The last two terms of the partition function can be ulteriorly worked out 

\begin{align*}
       &\hspace{-1cm}\mathbb{E}_{\hat{\tau}} \EXP{\hat{d}_{\hat\tau}  \hat{\tau}^2} \mathbb{E}_{\tilde s} \, \mathbb{E}_{\{\tau^{\gamma}\}} \exp{\left(   \frac{1}{2} \hat{q}_{\tau}  \left( \sum_{\gamma } \tau^{\gamma} \right)^2  -  \frac{1}{2} \hat{q}_{\tau} \sum_{ \gamma} (\tau^{\gamma})^2 + \hat{m}_{\tau}  \sum_{ \gamma}  \tau^{\gamma} \hat{\tau} + \hat{d}_{\tau} \sum_{ \gamma} (\tau^{\gamma})^2 +\tilde s \hat p_\tau \sum_\gamma  \tau^\gamma \right)}= \\
        =& \Bigg\langle \mathbb{E}_{\hat{\tau}}\, \EXP{\hat{d}_{\hat\tau} \hat{\tau}^2} \mathbb{E}_{\tilde s} \Bigg( \mathbb{E}_{\tau} \exp \left( -(\frac{\hat{q}_{\tau}}{2} -\hat{d}_{\tau} )\tau^2+\tau (\hat{m}_{\tau} \hat{\tau} +\sqrt{\hat{q}_{\tau}}z + \tilde s \,\hat{p}_\tau)  \right)  \Bigg)^n \Bigg\rangle_z  \\
        =&  \Bigg\langle \mathbb{E}_{\hat{\tau}}\, \EXP{\hat{d}_{\hat\tau} \hat{\tau}^2} \mathbb{E}_{\tilde s} \exp \left\{ n \log \Bigg[ \mathbb{E}_{\tau} \exp\left( -(\frac{\hat{q}_{\tau}}{2} -\hat{d}_{\tau} )\tau^2+\tau (\hat{m}_{\tau} \hat{\tau} +\sqrt{\hat{q}_{\tau}}z + \tilde s \,\hat{p}_\tau)  \right)  \Bigg]\right\} \Bigg\rangle_z 
        \\
        \simeq &
\Bigg\langle \mathbb{E}_{\hat{\tau}}\, \EXP{\hat{d}_{\hat\tau} \hat{\tau}^2} \mathbb{E}_{\tilde s} \left\{1+ n \log \Bigg[ \mathbb{E}_{\tau} \exp\left( -(\frac{\hat{q}_{\tau}}{2} -\hat{d}_{\tau} )\tau^2+\tau (\hat{m}_{\tau} \hat{\tau} +\sqrt{\hat{q}_{\tau}}z + \tilde s \,\hat{p}_\tau)  \right)  \Bigg] \right\} \Bigg\rangle_z 
\\
\simeq& \Bigg\langle \mathbb{E}_{\hat{\tau}}\, \EXP{\hat{d}_{\hat\tau} \hat{\tau}^2} + n \mathbb{E}_{\hat{\tau}}\, \EXP{\hat{d}_{\hat\tau} \hat{\tau}^2} \mathbb{E}_{\tilde s} \log \Bigg[ \mathbb{E}_{\tau} \exp\left( -(\frac{\hat{q}_{\tau}}{2} -\hat{d}_{\tau} )\tau^2+\tau (\hat{m}_{\tau} \hat{\tau} +\sqrt{\hat{q}_{\tau}}z + \tilde s \,\hat{p}_\tau)  \right)  \Bigg] \Bigg\rangle_z 
    \end{align*}
    \begin{equation*}
     \hspace{-1.2cm} \scalebox{0.98}{%
       $ \simeq \mathbb{E}_{\hat{\tau}}\, \EXP{\hat{d}_{\hat\tau} \hat{\tau}^2} \, \left\{1 + \frac{n}{\mathbb{E}_{\hat{\tau}}\, \EXP{\hat{d}_{\hat\tau} \hat{\tau}^2}}  \Bigg\langle\mathbb{E}_{\hat{\tau}}\, \EXP{\hat{d}_{\hat\tau} \hat{\tau}^2} \mathbb{E}_{\tilde s} \log \Bigg[ \mathbb{E}_{\tau} \exp\left( -(\frac{\hat{q}_{\tau}}{2} -\hat{d}_{\tau} )\tau^2+\tau (\hat{m}_{\tau} \hat{\tau} +\sqrt{\hat{q}_{\tau}}z + \tilde s \,\hat{p}_\tau)  \right)  \Bigg]\Bigg\rangle_z  \right\} 
     \,,$}
    \end{equation*}
    where we use the expansion of the exponential for small $n$. Reconstructing the term inside the free energy and substituting $\log(1+x) \simeq x$ for small $x$
    \begin{align*}
&\hspace{-2.2cm}\scalebox{0.98}{%
       $ \alpha \log \left(  \mathbb{E}_{\hat{\tau}}\, \EXP{\hat{d}_{\hat\tau} \hat{\tau}^2} \, \left\{1 + \frac{n}{\mathbb{E}_{\hat{\tau}}\, \EXP{\hat{d}_{\hat\tau} \hat{\tau}^2}}  \Bigg\langle\mathbb{E}_{\hat{\tau}}\, \EXP{\hat{d}_{\hat\tau} \hat{\tau}^2} \mathbb{E}_{\tilde s} \log  \mathbb{E}_{\tau} \exp\left( -(\frac{\hat{q}_{\tau}}{2} -\hat{d}_{\tau} )\tau^2+\tau (\hat{m}_{\tau} \hat{\tau} +\sqrt{\hat{q}_{\tau}}z + \tilde s \,\hat{p}_\tau)  \right)  \Bigg\rangle_z  \right\}  \right) \simeq $} \\
         & \hspace{-1cm}\simeq 
         \scalebox{0.98}{%
       $\,  \alpha \log \mathbb{E}_{\hat{\tau}}\, \EXP{\hat{d}_{\hat\tau} \hat{\tau}^2} + \frac{\alpha n}{\mathbb{E}_{\hat{\tau}} \,\EXP{\hat{d}_{\hat\tau} \hat{\tau}^2}} \Bigg\langle \mathbb{E}_{\hat{\tau}} \,\EXP{\hat{d}_{\hat\tau} \hat{\tau}^2}  \mathbb{E}_{\tilde s} \log \mathbb{E}_{\tau} \exp{\left( -(\frac{\hat{q}_{\tau}}{2} -\hat{d}_{\tau} )\tau^2+\tau (\hat{m}_{\tau} \hat{\tau} +\sqrt{\hat{q}_{\tau}}z +\tilde s \,\hat{p}_\tau)  \right)} \Bigg\rangle_z  
       $} \,,
    \end{align*}
the same procedure holds for 
\begin{align*}
    &\log \mathbb{E}_{\hat{{\xi}}} \EXP{\hat{d}_{\hat\xi} \hat{\xi}^2} \mathbb{E}_s \,\mathbb{E}_{\{{\xi}^{\gamma}\}} \exp{\left(   \frac{\hat{q}}{2} \bigg( \sum_{\gamma} \xi^{\gamma} \bigg)^2 -  \frac{\hat{q}}{2} \sum_{ \gamma} (\xi^{\gamma})^2 + \hat{m} \sum_{ \gamma} \xi^{\gamma} \hat{\xi} +  \hat{d} \sum_{ \gamma} (\xi^{\gamma})^2 + s\,\hat{p} \sum_\gamma \xi^\gamma \right)}\simeq \\
     & \simeq  \scalebox{0.98}{%
       $   \log \mathbb{E}_{\hat{\xi}} \EXP{\hat{d}_{\hat\xi} \,\hat{\xi}^2} + \frac{ n}{\mathbb{E}_{\hat{\xi}} \,\EXP{\hat{d}_{\hat\xi}\,\hat{\xi}^2}} \Bigg\langle \mathbb{E}_{\hat{\xi}} \,\EXP{\hat{d}_{\hat\xi}\, \hat{\xi}^2} \mathbb{E}_s  \log \mathbb{E}_{\xi} \exp{ \left( -(\frac{\hat{q}}{2} -\hat{d} )\xi^2+\xi (\hat{m}\, \hat{\xi} +\sqrt{\hat{q}}\,z+s\,\hat p)  \right)} \Bigg\rangle_z   
       $}\,.
\end{align*}
Collecting the previous two  expansions, and imposing the RS ansatz on the other terms of the free energy (\ref{eq:: Extr free energy}), the final extremization can be rewritten as \begin{equation*}
-\beta f (\beta,\hat{\beta},\gamma)
 \approx \lim_{ n\to 0} \frac{1}{ n} \extr_{\tiny \substack{\Lambda,\hat\Lambda\\
 \tiny \Lambda_\tau, \hat\Lambda_\tau }} \hat f^{RS}(\Lambda,\hat\Lambda,\Lambda_\tau, \hat\Lambda_\tau) \,,
\end{equation*} 
where now $\Lambda$ is the RS reduced set of the pattern-related order parameters $\Lambda =\{p,q,m,d\}$, while $ \hat\Lambda$ is the conjugated one ( $\Lambda_\tau, \hat\Lambda_\tau$ are the same for the hidden units). The free energy can be divided into two parts according to its power of $n$ 
\begin{align}
    &\hat f^{RS}(\Lambda,\hat\Lambda,\Lambda_\tau,\hat\Lambda_\tau)  = \hat f^{RS}_0 + n \hat f^{RS}_1 +\mathcal{O}(n^2) \,, \notag
\end{align}
\begin{align}
    \hat f^{RS}_0 &=  \frac{\alpha \hat{\beta}}{2} d_{\hat\tau} d_{\hat\xi} +\hat{\beta}\extr_{\bm p_T} f_T(\bm p_T) +\alpha \log \mathbb{E}_{\hat{\tau}} \EXP{\hat{d}_{\hat\tau} (\hat{\tau})^2}+ \log \mathbb{E}_{\hat{\xi}} \EXP{\hat{d}_{\hat\xi} (\hat{\xi})^2}-\alpha \hat{d}_{\hat\tau} d_{\hat\tau}-\hat{d}_{\hat\xi} d_{\hat\xi} \notag \;,\\
    \label{}
    \hat f^{RS}_1 &= \frac{1}{2}\hat{q}q+\alpha\frac{1}{2}\hat{q}_{\tau}q_{\tau}-\hat{m}m-\alpha \hat{m}_{\tau}m_{\tau}-\hat{d}d-\alpha \hat{d}_{\tau}d_{\tau}+\alpha^2 \frac{\beta}{2}\Omega_{\xi}p_{\text{\ensuremath{\tau}}}^{2}+\frac{\beta}{2}\Omega_{\tau}p^{2}+ \notag \\
    &+\frac{\alpha }{\mathbb{E}_{\hat{\tau}}\EXP{\hat{d}_{\hat\tau}\hat{\tau}^{2}}}\Bigg\langle\mathbb{E}_{\hat{\tau}}\EXP{\hat{d}_{\hat\tau}\hat{\tau}^{2}}\mathbb{E}_{s}\log\mathbb{E}_{\tau}\exp\Bigg(-(\frac{\hat{q}_{\tau}}{2}-\hat{d}_{\tau})\tau^{2}+\tau(\hat{m}_{\tau} \hat\tau+\sqrt{\hat{q}_{\tau}}z+\hat{p}_{\tau}s)\Bigg)\Bigg\rangle_{z}+ \notag\\    
    &+\frac{1}{\mathbb{E}_{\hat{\xi}}\EXP{\hat{d}_{\hat\xi}\hat{\xi}^{2}}}\Bigg\langle\mathbb{E}_{\hat{\xi}}\EXP{\hat{d}_{\hat\xi}\, \hat{\xi}^{2}}\mathbb{E}_{s}\log\mathbb{E}_{\xi}\exp\Bigg(-(\frac{\hat{q}}{2}-\hat{d})\xi^{2}+\xi(\hat{m}\hat\xi+\sqrt{\hat{q}}z+\hat{p}s)\Bigg)\Bigg\rangle_{z} + \notag \\
    & -\frac{\beta \alpha}{2}qq_{\tau}+\alpha \sqrt{\hat\beta \beta} m m_{\tau}-\alpha p_{\tau}\hat{p}_{\tau}-p\hat{p} +\frac{\beta \alpha}{2}d d_{\tau} +\beta\extr_{\bm p_S} f_S(\bm p_S)\,.\label{eq:free_energy_RS}
\end{align}
The extremization of $\hat f^{RS}_0$ involves only the teacher order parameters. The extremizers should be s.t. $\extr \hat f^{RS}_0=0$ in order to avoid divergences when $n\to0$. Indeed, remembering Eqs.(\ref{eq: A teacher den saddle point}), the parameters of order zero in $n$ satisfy the following
\begin{align}
\label{Aeq:: n^0 order params}
    \hat{d}_{\hat\xi} &= 0\,, \quad d_{\hat\xi}=1\,,\\
    \hat{d}_{\hat\tau} &=\frac{\hat\beta}{2}\,, \quad d_{\hat\tau}= \langle \hat\tau ^2\rangle_{\hat\tau} \notag \,,
\end{align}
which implies $\hat f^{RS}_0=0$ and ensure the student partition function is not diverging. Using Eqs.(\ref{A student denominator}) and updating $\hat f_1^{RS}$ with (\ref{Aeq:: n^0 order params}) we obtain the values of $\Lambda$ and $\Lambda_{\tau}$ that extremize $\hat f_1^{RS}$
\begin{align}
      \hat{p}_{\tau} &= \beta \alpha \Omega_{\xi} p_{\tau}\,, \quad \hat{m}_{\tau} = \sqrt{\beta \hat{\beta}} m \,,
    \quad \hat{q}_{\tau} = \beta q\,, \quad \hat{d}_{\tau} = \beta \frac{d}{2}\,, \\
    \hat{p} &= \beta \Omega_{\tau} p\,, \quad \hat{m}= \alpha \sqrt{\beta \hat{\beta}} m_{\tau}\,, \quad \hat{q} = \alpha \beta q_{\tau}\,, \quad \hat{d} = \alpha \beta \left(\frac{d_{\tau}}{2} -\langle \tau^2 \rangle_\tau \right)\,,
\end{align}
\begin{gather}
\label{Aeq:: MF eqs}
p = \big\langle s\langle \xi  \rangle_{\xi|z,s,\hat{\xi}} \big\rangle_{z,s,\hat{\xi}}\qquad \qquad
p_{\tau} = \big\langle s \langle \tau  \rangle_{\tau|z,s,\hat{\tau}} \big\rangle_{z,s,\hat{\tau}}
\\
m = \big\langle \hat{\xi} \langle \xi   \rangle_{\xi|z,s,\hat{\xi}} \big\rangle_{z,s,\hat{\xi}}
\qquad \qquad
m_{\tau} = \big\langle \hat{\tau} \langle \tau   \rangle_{\tau|z,s,\hat{\tau}} \big\rangle_{z,s,\hat{\tau}}
\\
q = \big\langle \langle \xi^{2} \rangle_{\xi|z,s,\hat{\xi}} \big\rangle_{z,s,\hat{\xi}}
\qquad\qquad
q_{\tau} = \big\langle \langle \tau  \rangle_{\tau|z,s,\hat{\tau}}^2 \big\rangle_{z,s,\hat{\tau}}
\\
d = \big\langle \langle \xi^2  \rangle_{\xi|z,s,\hat{\xi}} \big\rangle_{z,s,\hat{\xi}}
\qquad\qquad
d_{\tau} = \big\langle \langle \tau^2   \rangle_{\tau|z,s,\hat\tau} \big\rangle_{z,s,\hat\tau} \,.
\label{Aeq:: MF self overlap}
\end{gather}
The internal distributions that average $\xi$ and $\tau$ are respectively 
\begin{align}
\label{Aeq::MFdistributions1}
    P(\xi|z,s,\hat{\xi})=&P_\xi (\xi)\exp\left( -\left(\frac{\hat{q}}{2}-\hat{d}\right)\xi^{2}+\xi\left(\hat{m}\hat{\xi}+\sqrt{\hat{q}}z+\hat{p}s\right)\right)\,,\\
    P(\tau|z,s,\hat{\tau})=&P_\tau(\tau)\exp\left( -\left(\frac{\hat{q}_{\tau}}{2}-\hat{d_{\tau}}\right)\tau^{2}+\tau\left(\hat{m_{\tau}}\hat{\tau}+\sqrt{\hat{q_{\tau}}}z+\hat{p_{\tau}}s\right)\right)\,,
\label{Aeq::MFdistributions2}
\end{align}
while the outher distributions $P_{\hat{\xi}}(\hat{\xi}) $ and $P_{\hat{\tau}}(\hat{\tau})\EXP{\frac{\hat{\beta}}{2}(\hat{\tau})^2} $ depend from the choice of the planted configuration and $z$ is a Normal random variable. 

%% file: Explicit_Generic_Eqs.tex
\section{RS equations in the interpolating case}
\label{Sec:: Appendix B}
So far, Eqs. (\ref{Aeq:: MF eqs}-\ref{Aeq:: MF self overlap}) are general and valid for any arbitrary choice of the priors for all variables.
In this section we derive explicitly the saddle point equations, replacing each of the random variables with their prior interpolation (\ref{eq:: x interpol}). The expressions of the distributions (\ref{Aeq::MFdistributions1}) and (\ref{Aeq::MFdistributions2}) can be recast as follows:
\begin{align}
    Z_{\xi}&^{-1}\sum_{\epsilon} \EXP{-\frac{\xi^2}{2 \gamma}+ \xi\left(\phi \epsilon +h\right)}\,,\notag \\
     Z_{\tau}&^{-1}\sum_{\epsilon_{\tau}} \EXP{-\frac{\tau^2}{2 \gamma_{\tau}}+ \xi\left(\phi_{\tau} \epsilon_{\tau} +h_{\tau}\right)}\,,\notag
\end{align}
where we use the shorthand notation 
\begin{align}
    \gamma = \frac{\Omega_{\xi}}{1-\alpha \beta \Omega_{\xi} \left(d_{\tau}-q_{\tau}-\langle \tau^2 \rangle_\tau \right)},\; 
\quad & \quad
\gamma_{\tau} = \frac{\Omega_{\tau}}{1- \beta \Omega_{\tau} \left(d -q  \right) }\,, \notag \\
\phi =  \frac{\sqrt{\delta}}{\Omega_{\xi}} \,, \quad & \quad \phi_{\tau} =  \frac{\sqrt{\delta_{\tau}}}{\Omega_{\tau}}  \,, \notag\\
h = \hat{m}\hat{\xi}+\sqrt{\hat{q}}z+\hat{p}s\,, \quad & \quad h_{\tau} = \hat{m}_{\tau}\hat{\tau}+\sqrt{\hat{q}_{\tau}}z+\hat{p}_{\tau}s\, \notag.
\end{align}
The averages in the mean field equations (\ref{Aeq:: MF eqs}-\ref{Aeq:: MF self overlap}) become:
\begin{align*}
    \langle \xi  \rangle_{\xi|z,s,\hat{\xi}} =& \partial_h \ln Z_{\xi} = \frac{\sum_{\epsilon} \EXP{ \frac{\gamma}{2} \left(\phi \epsilon +h\right)^2 } \gamma \left(\phi \epsilon +h\right)}{\sum_{\epsilon} \EXP{ \frac{\gamma}{2} \left(\phi \epsilon +h\right)^2 }} 
    = \gamma \,h + \gamma \phi \tanh{\left(\gamma \phi \, h \right)}\\
     \langle \xi^2  \rangle_{\xi |z,s,\hat{\xi}} =& \partial_h ^2 \ln Z_{\xi} + \langle \xi  \rangle_{\xi|z,s,\hat{\xi}} ^2 
     =\gamma + \gamma^2 (h^2 + \phi^2 ) +2 \gamma^2 \phi \,h \tanh{\left( \gamma \phi \,h \right)}\,,
\end{align*}
the same is also valid for the corresponding $\tau$ counterparts. We can also proceed with the evaluation of the averages on the planted teacher and the example randomness, using their interpolating decomposition\begin{equation*}
    \mathbb{E}_{s,\hat{\xi}}f(s,\hat{\xi}) = \mathbb{E}_{g_{s,\hat{\xi}}}\mathbb{E}_{\epsilon_{s,\hat{\xi}}}f(\sqrt{\Omega_{s}}\,g_{s} + \sqrt{\delta_{s}}\epsilon_{s},\sqrt{\Omega_{\hat\xi}}\,g_{\hat\xi} + \sqrt{\delta_{\hat\xi}}\,\epsilon_{\hat\xi})\;.
\end{equation*}
This leads to the final form 
\begin{align}
    \label{eq:: explicit form equations xi}p=&\gamma\beta_{ } \Omega_{\tau}p+\gamma^{2}\phi^{2}\Omega_{s}\beta_{ } \Omega_{\tau}p\left(1-\bar{q}\right)+\gamma\phi\sqrt{\delta_{s}}\bar{m_{s}},\\ 
m=&\gamma\alpha\sqrt{\hat{\beta}\beta }m_{\tau}+\gamma^{2}\phi^{2}\Omega_{\hat\xi}\,\alpha\sqrt{\hat{\beta}\beta_{ } }\,m_{\tau}\left(1-\bar{q}\right)+\gamma\phi\sqrt{\delta_{\hat\xi}}\,\bar{m}_{*}\\
q=&\gamma^{2}\left(\left(\alpha\sqrt{\hat{\beta}\beta_{ } }\,m_{\tau}\right)^{2}+\alpha\beta_{ } q_{\tau}+(\beta_{ } \Omega_{\tau}p)^{2}\right)+\gamma^{2}\phi^{2}\bar{q}+\\
+&2\gamma^{2}\phi\left(\beta_{ } \Omega_{\tau}p\sqrt{\delta_{s}}\bar{m_{s}}+\alpha\sqrt{\hat{\beta}\beta_{ } }q_{\tau}\sqrt{\delta_{\hat\xi}}\,\bar{m}_{*}\right)+ \notag \\
+&2\gamma^{3}\phi^{2}\left((\beta_{ } \Omega_{\tau}p)^{2}\Omega_{s}+(\alpha\sqrt{\hat{\beta}\beta_{ } }\,m_{\tau})^{2}\Omega_{\hat\xi}+\alpha\beta_{ } q_{\tau}\right)\left(1-\bar{q}\right) \notag \\
\label{eq:: self overlap}
d=&\gamma+\gamma^{2}\phi^{2}\left(1-\bar{q}\right)+q\;.
\end{align}
Here we can distinguish the effective magnetization related to the signal and the example, along with an effective overlap
\begin{gather}
\label{eq:: effective magnetization}
    \bar{m}_{*}=  \left\langle \mathbb{E}_{s}\tanh\left( \gamma \phi \left(\alpha m_{\tau}\sqrt{\hat{\beta}\beta\delta_{\hat\xi}}+\sigma z+\beta_{ } \Omega_{\tau}p s\right)\right)\right\rangle_z  \\
    \bar{m}_{s}= \left\langle \mathbb{E}_{s} \epsilon_s \tanh\left( \gamma \phi \left(\alpha m_{\tau}\sqrt{\hat{\beta}\beta\delta_{\hat\xi}}+\sigma z+\beta_{ } \Omega_{\tau}p s\right)\right)\right\rangle_z \notag \\
    \bar{q}= \left\langle \mathbb{E}_{s}\tanh^2\left( \gamma \phi \left(\alpha m_{\tau}\sqrt{\hat{\beta}\beta \delta_{\hat\xi}}+\sigma z+\beta_{ } \Omega_{\tau}p s\right)\right)\right\rangle_z \notag \;.
\end{gather}
where $\sigma^2 = \left(\alpha q_{\tau}\sqrt{\hat{\beta}\beta\Omega_{\hat\xi}}\,\right)^2 + \alpha \beta q_{\tau}$. The same can be done for the $\tau$ side. This turns out not to be a completely symmetric case,  due to the presence of the Gaussian factor inside the average over $\hat{\tau}$
\begin{equation*}
    \mathbb{E}_{\hat{\tau}}\,\EXP{\frac{\hat{\beta}}{2}(\hat{\tau})^{2} } f(\hat{\tau})= \mathbb{E}_{\hat{g}_{\tau}}\mathbb{E}_{\hat{\epsilon}_{\tau}}\,\EXP{(\sqrt{\Omega_{\hat\tau}}\,g_{\hat\tau} + \sqrt{\delta_{\hat\tau}}\,\epsilon_{\hat\tau})^2}f(\sqrt{\Omega_{\hat\tau}}\,g_{\hat\tau} + \sqrt{\delta_{\hat\tau}}\,\epsilon_{\hat\tau})\,,
\end{equation*}
this difference is due to the presence of the self spherical constraint (\ref{eq: A aprox teacher z 1}) in the teacher model. The hidden set of equations are
\begin{align}
    \label{eq:: explicit form equations tau}
    p_{\tau}=&\gamma_{\tau}\beta_{ } \alpha\Omega_{\xi} p_{\tau}+\gamma_{\tau}^{2}\phi_{\tau}^{2}\Omega_{s} \beta_{ } \alpha\Omega_{\xi}\, p_{\tau}\left(1-\bar{q}_{\tau}\right)+\gamma_{\tau}\phi_{\tau}\sqrt{\delta_{s}}\bar{m}_{\tau,s}\\
    m_{\tau}=&\gamma_{\tau}\sqrt{\hat{\beta}\beta }\,m \left(\frac{\delta_{\hat\tau} + \Omega_{\hat\tau}(1-\hat{\beta}\Omega_{\hat\tau})}{(1-\hat{\beta}\Omega_{\hat\tau})^{2}}\right)+\frac{\sqrt{\delta_{\hat\tau}}}{\left(1-\hat{\beta}\Omega_{\hat\tau}\right)}\gamma_{\tau}\phi_{\tau}\bar{m}_{\tau,*}\\
    +&\frac{\gamma_{\tau}^{2}\phi_{\tau}^{2}}{(1-\hat{\beta}\Omega_{\hat\tau})}\sqrt{\hat{\beta}\beta_{ } }\,m \,\Omega_{\hat\tau}\left(1-\bar{q}_{\tau}\right)\\
    q_{\tau}=&\gamma_{\tau}^{2}\left[ \beta_{ } q +(\beta_{ } \alpha\Omega_{\xi}\, p_{\tau})^{2}+(\sqrt{\hat{\beta}\beta_{ } }\,m)^{2}\left(\frac{\delta_{\hat\tau}+\Omega_{\hat\tau}(1-\hat{\beta}\Omega_{\hat\tau})}{(1-\hat{\beta}\Omega_{\hat\tau})^{2}}\right)\right] + \gamma_{\tau}^{2}\phi_{\tau}^{2}\bar{q}_{\tau}+\\
+&2\gamma_{\tau}^{2}\phi_{\tau}\left(\beta_{ } \alpha\Omega_{\xi} p_{\tau}\sqrt{\delta_{s}}\bar{m}_{\tau,s}+\frac{\sqrt{\hat{\beta}\beta_{ } }\,m\sqrt{\delta_{\hat\tau}}}{(1-\hat{\beta}\Omega_{\hat\tau})}\bar{m}_{\tau,*}\right) \notag\\ 
    +&2\gamma_{\tau}^{3}\phi_{\tau}^{2}\left(\Omega_{s} (\beta_{ } \alpha\Omega_{\xi} p_{\tau})^{2}+\beta_{ } q+\frac{(\sqrt{\hat{\beta}\beta_{ } }\,m)^{2}\Omega_{\hat\tau}}{(1-\hat{\beta}\Omega_{\hat\tau})}\right)\left(1-\bar{q}_{\tau}\right) \notag\\
    \label{eq:: self overlap tau}
    d_{\tau}=&\gamma_{\tau}^{2}\phi_{\tau}^{2}\left(1-\bar{q}_{\tau}\right)+\gamma_{\tau}+q_{\tau} \;.
\end{align}
The effective magnetizations and overlap in the $\tau$ version are
\begin{align}
\label{eq:: effective magn TAU}
   \bar{m}_{\tau, *}&= \frac{1}{\mathbb{E}_{\hat{\tau}}\,\EXP{\frac{\hat\beta}{2}(\hat{\tau})^{2}}}\left\langle \mathbb{E}_{s}\,\mathbb{E}_{\hat{\tau}}\,\EXP{\frac{\hat\beta}{2}(\hat{\tau})^{2}}\epsilon_{\hat\tau}\tanh\left(\gamma_\tau \phi_\tau \left(\sqrt{\hat\beta \,\beta}\,m\hat{\tau}+\sqrt{\beta q}z+\beta\alpha\Omega_{\xi} p_{\tau}s\right)\right)\right\rangle _{z} \\
   \bar{m}_{\tau, s}&= \frac{1}{\mathbb{E}_{\hat{\tau}}\EXP{\frac{\hat\beta}{2}(\hat{\tau})^{2}}}\left\langle \mathbb{E}_{s}\,\mathbb{E}_{\hat{\tau}}\EXP{\frac{\hat\beta}{2}(\hat{\tau})^{2}}\epsilon_{\tau,s}\tanh\left(\gamma_\tau \phi_\tau \left(\sqrt{\hat\beta \, \beta}\,m\hat{\tau}+\sqrt{\beta q}z+\beta_{2}\alpha\Omega_{\xi} p_{\tau}s\right)\right)\right\rangle _{z} \notag\\
   \bar{q}=&  \frac{1}{\mathbb{E}_{\hat{\tau}}\EXP{\frac{\hat\beta}{2}(\hat{\tau})^{2}}}\left\langle \mathbb{E}_{s}\,\mathbb{E}_{\hat{\tau}}\EXP{\frac{\hat\beta}{2}(\hat{\tau})^{2}}\tanh^2 \left(\gamma_\tau \phi_\tau \left(\sqrt{\hat\beta \,\beta}\,m\hat{\tau}+\sqrt{\beta q}z+\beta \alpha\Omega_{\xi} p_{\tau}s\right)\right)\right\rangle _{z} \notag \,.
\end{align}